\documentclass{cimento}

\usepackage{graphicx}
\usepackage{dcolumn}
\usepackage{bm}
\usepackage{amsmath}
\usepackage{amssymb}
\usepackage{float}
\usepackage[titletoc,title]{appendix}

\title{Many-Body Physics and Quantum Simulations with Strongly Interacting Photons}
\author{Jirawat~Tangpanitanon\from{ins:x}\ETC,
        \atque
Dimitris G.~Angelakis\from{ins:x},\from{ins:y}}
\instlist{\inst{ins:x} Centre for Quantum Technologies, National University of Singapore, 3 Science Drive 2, Singapore 117543
\inst{ins:y} School of Electrical and Computer Engineering, Technical University of Crete, Chania, Greece 73100}

\begin{document}

\maketitle

\begin{abstract}
Simulating quantum many-body systems on a classical computer generally requires a computational cost that grows exponential with the number of particles. This computational complexity has been the main obstacle to understanding various fundamental emergent phenomena in condensed matters such as high-Tc superconductivity and the fractional quantum-Hall effect. The difficulty arises because even the simplest models that are proposed to capture those phenomena cannot be simulated on a classical computer. Recognizing this problem in 1981, Richard Feynman envisioned a quantum simulator, an entirely new type of machine that exploits quantum superposition and operates by individually manipulating its constituting quantum particles and their interactions. Recent advances in various experimental platforms from cold atoms in optical lattices, trapped ions, to solid-state systems have brought the idea of Feynman to the realm of reality.  Among those, interacting photons in superconducting circuits has been one of the promising platforms thanks to their local controllability and long coherence times. Early theoretical proposals have shown possibilities to realize quantum many-body phenomena of light using coupled cavity arrays such as Mott to superfluid transitions and fractional quantum Hall states. Start-of-the-art experiments include realization of interacting chiral edge states and stroboscopic signatures of localization of interacting photons in a three-site and a nine-site superconducting circuit, respectively. Interacting photons also serve as a natural platform to simulate driven-dissipative quantum many-body phenomena. A 72-site superconducting circuit has also recently been fabricated to study a dissipative phase transition of light.
\end{abstract}


\section{Introduction}

Quantum simulation is an emerging interdisciplinary field in physics \cite{2012_zoller_natphy, 2012_lewenstein_rpp,2014_dieter_epj}. It aims to develop a new type of devices that exploit quantum coherence to answer questions about models that describe complex quantum phenomena that are beyond the reach of a classical computer. Experimental progress in the past 30 years have made it possible to control and manipulate individual quantum systems including trapped ions \cite{2012_blatt_np}, cold atoms in optical lattices \cite{2017_gross_sci}, nuclear magnetic resonance (NMR) \cite{2017_andrea_natcomm}, interacting photons \cite{2014_lukin_natphy}, quantum dots \cite{2017_vadersypen_nat}, superconducting circuits \cite{2012_koch_natphy}, and nitrogen-vacancy centers \cite{2012_lukin_natcom}. These new exciting developments transit quantum simulation from a theoretical proposal to the realm of reality. 

Below, we first discuss the concept of simulation on a classical computer, its limitation, and motivation of quantum simulation. We then review various experimental platforms for quantum simulation. We conclude the section by giving an overview of the lecture which will be focused on interacting photons in superconducting circuits for quantum simulation of both in and out-of-equilibrium quantum many-body systems. 

\subsection{Computer simulation}
\label{sec:quantum_simulation}

Imitating a complex real-world process or system by simulating relevant models on a classical computer has been an essential technique for the development of science and technology. To simulate something, one needs first to develop a model that describes characteristics and behavior of such system \cite{7acdd06b7ccc467faa7a4676ae3726cc}. By changing variables of the model in the simulator, one can make predictions about the behavior of the real complex system. In many cases, these predictions can be used to reduce a high cost of performing several trial-and-error experiments on the real system. For example, the drug design process can be drastically speed up by appropriately modeling molecular systems \cite{drug}. An airplane wing can be designed by simulating relevant drag forces via fluid dynamics models \cite{VALLESPIN20122}. 

In some cases, simulation is useful when it is difficult or not possible to perform experiments on a real system. For example, the climate system can be predicted by solving differential equations that represent essential factors of climate and their interactions including atmosphere, oceans, land surface, and ice \cite{Gilchrist}. We note that the act of simulation itself, \textit{i.e.}, imitating real-world processes or systems, is not necessarily done on a classical computer. For example, simulating a weightless in the outer space can be done by aerobatic maneuvers on the Earth that undergo a parabolic motion \cite{plane}.

\subsection{Quantum simulation}

Despite the tremendous success of computer simulation over the past 70 years from the nuclear detonation process in the Manhattan Project in World War II \cite{firstcom} to forecasting of prices on financial markets \cite{2006_seydel}, there remains a large class of systems that are too complex to be simulated by any conceivable classical computer. In physics, this usually involves simulating systems that are non-linear or chaotic due to their non-integrability. In quantum mechanics, although the Schr\"{o}dinger equation is a linear equation, simulating it generally requires a computational cost that grows exponential with the number of particles. For example, to describe a wavefunction of $N$ spin-$1/2$ particles, one needs to store $2^N$ complex coefficients. Also, one needs to perform linear algebra to such vectors in order to evaluate physical observables that describe the dynamics or even the ground state of the system. This task can be impossible when $N$ is as small as 50 which requires several petabytes of classical memories \cite{2017_steiger_arxiv,2017_wisnieff_arxiv}. This number is far less than the number of electrons in real materials which are of the order of $10^{23}$. This computational complexity is the main obstacle to understanding various fundamental emergent phenomena such as high-$T_c$ superconductivity \cite{0034-4885-71-1-012501} and the fractional quantum-Hall effect \cite{RevModPhys.89.025005}.

Recognizing this problem, in 1981 Richard Feynman envisioned an idea of a quantum simulator, a machine that exploits quantum parallelism and operates by individually manipulating its constituting quantum particles and their interactions \cite{1982_feynman_ijtp}. Predictions are made by performing appropriate measurements on those particles. Feynman proposed to quantize both space and time to allow such simulator to be universal, \textit{i.e.}, can be programmed to simulate any quantum systems. The idea was later extended by Seth Lloyd in 1996 \cite{1996_lloyd_sci}, who proved that by evolving in small time steps, or trotterization, such simulator could simulate the dynamics of any local quantum many-body Hamiltonian with the time scale that grows only polynomially with the number of particles. 

Such universal or a `digital' quantum simulator, however, requires full control over quantum many-body systems and may still be a long time ahead. Alternatively, one may aim at a less ambitious goal of an `analog' quantum simulator. The idea is to use reasonably well-controlled quantum systems to simulate only certain classes of quantum systems which are, nevertheless, interesting and cannot be simulated on a classical computer. Building the later may be less prone to errors because it does not requite trotterization. Also, phases of matter are typically robust against local perturbation. Nevertheless, as pointed out in ref.\cite{2012_lewenstein_rpp}, a functioning quantum simulator should (i) be able to mimic a simple model, or a family of simple models, (ii) simulate models that are of some relevance for applications, (iii) simulate models that are computationally hard for classical computers, and (iv) allow for broad control of the parameters of the simulated model. Also, a quantum simulator should allow for validation, for example, by benchmarking against a classical computer in the regimes where numerical or analytical techniques exist or against different quantum simulators, ideally implemented in different platforms which subjected to different noises.

It might be hard, if not impossible, to prove that a given system cannot be efficiently simulated with a classical computer. Many quantum many-body systems can be simulated on a classical computer with approximate numerical methods such as artificial neural networks \cite{2017_troyer_sci}, tensor networks \cite{Orus}, dynamical mean-field theory, density matrix renormalization group (DMRG) theory \cite{schollwoeck2011}, density functional theory \cite{10.1007/978-94-009-9027-2_2} and quantum Monte-Carlo \cite{VONDERLINDEN199253}. However, they are known to be limited to certain classes of problems. For example, DMRG is appreciable only to gapped systems in one dimension. Quantum Monte Carlo does not work with fermionic statistics or frustrated models, due to the sign problem. Mean-field theory only works when the correlation between sites is weak and often fails in one dimension. 

With this in mind, the models that benefited most from a quantum simulator are expected to be the ones that involve a large amount of entanglements such as zero-temperature ground states of many-body Hamiltonians near the phase transition, non-equilibrium dynamics of driven or quenched systems, and dissipative dynamics of open systems. A quantum simulator can, for example, rule out or validate candidate models such as the Femi-Hubbard model for describing high-temperature superconductivity \cite{2006_leggett_natphy}, check the eigenstate thermalization hypothesis with various quantum many-body systems \cite{2015_huse_2015}, and compute accurate calculations of molecular properties for quantum chemistry \cite{2011_guzik_arpc,2010_white_natchem}. 

\subsection{Platforms for quantum simulation}
\subsubsection{\textbf{Cold neutral atoms in optical lattices:}}

Ultracold atoms in optical lattices represent one of the most versatile platforms for quantum simulation  \cite{2017_gross_sci,2012_sylvian_natphy,2008_zwerger_aps}. Optical lattices are formed by interfering laser beams in different directions to create a controllable standing-wave pattern that mimics the crystal lattice of a solid. Atoms can be trapped in the optical lattice due to an effective periodic potential landscape induced by laser beams via a dipole moment of the atoms. Ultracold atoms in optical lattices were first used to simulate the celebrated Mott to the superfluid phase transition in the Bose-Hubbard model \cite{2001_bloch_nat}. Subsequent work has shown quantum gas microscopes which enable fluorescence detection of atoms in single sites \cite{2010_stefan_nat}, quantum magnetism \cite{Lewenstein292}, possibilities to create artificial gauge fields by lattice shaking or by laser-induced tunneling \cite{2016_zoller_natphy}, and realization of the Fermi-Hubbard model \cite{PhysRevLett.114.193001, 2015_kuhr_natphy,PhysRevLett.114.213002, PhysRevLett.115.263001}. The Bose-Einstein condensation to Bardeen-Cooper-Schrieffer crossover was also observed in the continuum limit \cite{bec}. Recently, cold atoms in optical lattices were recently used to study the breakdown of thermodynamics description of interacting boson gas in two-dimensional disordered lattices \cite{2016_gross_sci}. Predicting a thermalized to a many-body localized phase transition in such a system is currently not possible with a classical computer due to the lack of efficient numerical techniques. A quantum simulator with 51-cold atoms trapped using optical tweezers has also been realized to observe different ordering in quantum Ising model \cite{2017_lukin_nat}. 
 
\subsubsection{\textbf{Trapped ions:}}
 
Another approach for atom-based quantum simulators is the use of trapped atomic ions held in linear radio-frequency traps \cite{2012_roos_natphy, PhysRevLett.92.207901}. Here ion crystals are formed by balancing the Coulomb repulsion between ions and trap confinement force, allowing them to be accurately controlled and manipulated. A wide range of models have been simulated in trapped ion systems from spin models \cite{2011_monroe_natcomm,2010_monroe_prb}, to dynamical phase transitions \cite{2017_monroe_nat} and discrete time crystal \cite{2017_monroe_nat}. High controllability in trapped ions also makes it a promising platform for quantum computing \cite{HAFFNER2008155, 1367-2630-15-12-123012}. The number of ions in a quantum simulator varies by a large factor depending on their controllability. For example, in 2012 a few hundreds of trapped ions with no local control was used to realize the quantum Ising model \cite{2012_bollinger_nat}, while a fully-programmable quantum simulator was only recently realized with five atoms in 2016 \cite{2016_monroe_nat}. This controllability is also a crucial factor for building a scalable quantum simulator in addition to the number of constituting atoms. Building a scalable quantum processor with high controllability and long coherent time over a few hundred qubits and defining relevant real-world applications are near-term challenges faced by all quantum technologies platforms \cite{2058-9565-3-3-030201}.

\subsubsection{\textbf{Solid-state systems:}}

There are also platforms for quantum simulation that based on solid state systems. For example, nitrogen-vacancy centers in diamond have been recently used to observe signatures of discrete time crystal \cite{2017_lukin_nat}. Donor spins in silicon have been used together with Nuclear Magnetic Resonance technique to demonstrate quantum gates between two qubits \cite{2017_andrea_natcomm}. A programmable quantum processor consisting of two single-electron-spin qubits in a silicon/silicon germanium (Si/SiGe) double quantum dot has been illustrated \cite{2018_vandersypen_nat}. The Fermi-Hubbard model has also been simulated using a quantum dot array in a GaAs/AlGaAs heterostructure semiconductor \cite{2017_vadersypen_nat}.

\subsubsection{\textbf{Interacting photons:}}

In parallel to the above progress, a new type of quantum simulators based on photons and hybrid light-matter excitations, known as polaritons, has been slowly emerging \cite{2017_angelakis_rpp, 2017_angelakis_springer}, inspired by advances in the field of quantum nonlinear optics and cavity quantum electrodynamics (QED) in the last two decades \cite{2014_lukin_natphy}. Pioneer theoretical works have shown possibilities to realize strongly correlated states of lights in coupled resonator arrays (CRAs) and to observe the Mott to the superfluid phase transition of light \cite{2007_dimitris_pra,2006_hartmann_natphy,2006_greentree_natphy}. Subsequent works extend the results to a family of many-body phenomena including an artificial field for the fractional quantum Hall effect \cite{2008_angelakis_prl, 2011_girvin_njp,2016_roushan_natphy}, effective spin models \cite{2007_hartmann_prl,2008_angelakis_el, 2009_li_ep,2012_sarkar_pb}, and topological transport of quantum states \cite{2016_tangpanitanon_prl}. Signatures of localization of interacting photons in a quasi-periodic potential have recently been observed with a nine-site superconducting circuit by directly measuring statistics of eigenenergies and spreading of energy eigenstates \cite{2018_tangpanitanon_sci}. This platform will be the main focus of this lecture.

In complementary with cold atoms, interacting photons provide a natural setting for simulating open quantum systems because light-matter systems dissipate to the environment and because they can be driven by external fields. The coupling to the environment is usually weak, and the bath is memoryless. Consequently, the system may reach a dynamically-stable steady state that depends on the symmetries of the system \cite{PhysRevA.89.022118}. Early theoretical works have shown that such steady states manifest various quantum many-body phases \cite{2009_carusotto_prl, 2010_hartmann_prl, 2012_dimitris_njp,2012_carusotto_prl, 2012_bardyn_prl, 2013_jin_prl} and can exhibit dissipative phase transitions (DPT) \cite{PhysRevA.86.012116, doi:10.1063/1.4978328}. A 72-site nonlinear superconducting circuit has recently been fabricated to study DPT with light \cite{2016_houck_prx}.

Perhaps, the most promising platform for realizing interacting photons is superconducting circuits where conventional optical and electron-beam lithography is used, allowing CRAs to be designed with great flexibility and high controllability \cite{2012_koch_natphy}. The circuit is made superconducting by cooling to few milli-Kelvins using a dilution refrigerator. Photonic modes can be realized from the co-planar transmission line or an LC circuit which effectively acts as a Fabry-Perot microwave cavity \cite{2012_houck_pra}.  An `artificial' two-level atom can be made from the use of Josephson junctions \cite{2005_nori_physto}. Both strong coupling \cite{2004_wallraff_nat} and ultra-strong coupling \cite{2016_kouichi_natphy} between an artificial atom and transmission line have also been reported. Non-linear coupled resonator arrays up to 19 sites have been implemented using superconducting circuits \cite{2018_tangpanitanon_sci,2017_ibm_nat,2017_regetti_arxiv}.

We note that there is also active research in the field of exciton polaritons in semiconductor materials, realizing quantum fluid of light \cite{2013_ciuti_rmp}. However, the interaction strength of such a system is typically weak at a few-photon level. Nevertheless, there are possibilities to enhance such interaction, for example, by resonantly coupling a pair of cavity polaritons to a biexciton state \cite{0295-5075-90-3-37001}. Experiments in this platform have led to realization of exotic phases of matter such as non-equilibrium Bose-Einstein condensation \cite{2006_dang_nat, 2013_rainer_natmat} and non-equilibrium polariton superfluidity \cite{2008_deveaud_natphy, 2009_alberto_natphy}.

\subsubsection{\textbf{conclusions:}}

In this section, we have discussed several platforms for quantum simulators. Next, we will discuss the basic concept of quantum phase transition with a specific example of the Mott to the superfluid phase transition in the Bose-Hubbard model. We then discuss the basic concepts in light-matter interaction, including field quantization in a cavity QED, the Jaynes-Cummings model, and photon blockade. We then review the early proposal for Mott to superfluid transitions of light, state-of-the-art experiments, and various works on both equilibrium and driven-dissipative many-body phases of light in CRAs. Lastly, we discuss circuit quantization and recent experimental progress in achieving interacting photons in superconducting circuits.


\section{Quantum phase transitions}

\begin{figure}
\centering
  \includegraphics[width=0.8\textwidth]{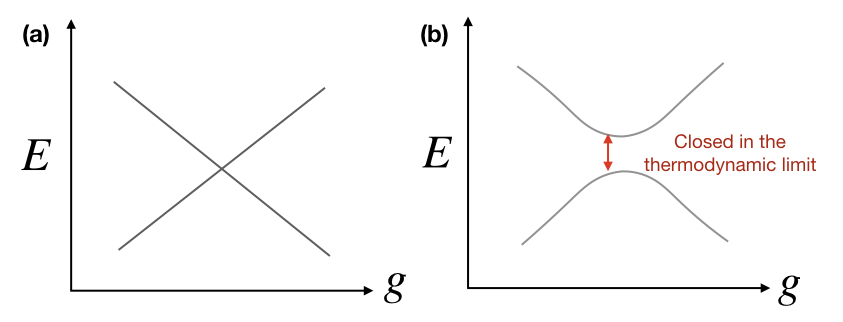}
\caption{\textbf{Two types of non-analyticity at a quantum phase transition.} Eigenenergies of $\hat{H}=\hat{H}_1+g\hat{H}_2$ as a function of $g$ in the case of \textbf{(a)} $\left[\hat{H}_1,\hat{H}_2\right]=0$ and \textbf{(b)}  $\left[\hat{H}_1,\hat{H}_2\right]\neq 0$. }
\label{fig:qpt}
\end{figure}

Identifying phases of matter are one of the main goals in condensed matter and material science. During a phase transition, specific properties of the material change abruptly as a result of the change of some external parameters. In classical physics, these parameters could be, for example, temperature, pressure, electric or magnetic fields. Classical phase transitions are driven by thermal fluctuations and cease to exist at zero temperature. Quantum phase transitions (QPTs), on the other hand, exist at zero temperature and are driven by quantum fluctuations \cite{2011-Sachdev}. Although, strictly speaking, absolute zero temperature is not physically realizable, signatures of QPTs can be observed when the energy scale of the thermal fluctuation $k_BT$ is much smaller than that of the quantum fluctuations $\hbar \omega$, where $\omega$ is a typical frequency of the quantum oscillation. 

To concretize the above description of QPT, let us consider a Hamiltonian of the form
\begin{equation}
\hat{H}=\hat{H}_1+g\hat{H}_2,
\nonumber
\end{equation}
where $g$ is a dimensionless parameter. The QTP concerns with non-analytic dependence of the ground-state energy 
\begin{equation}
E(g)=\langle G |\hat{H}|G\rangle
\nonumber
\end{equation}
as the parameter $g$ changed. Here $|G\rangle$ is the ground state of the system, \textit{i.e.}, $\hat{H}|G\rangle=E(g)|G\rangle$. In the case of $\left[\hat{H}_1,\hat{H}_2\right]=0$, non-analyticity can happen due to crossing of eigenvalues, see fig. (\ref{fig:qpt}). In the case of $\left[\hat{H}_1,\hat{H}_2\right]\neq 0$, non-analyticity can happen due to the closing of the energy gap between the ground state and the first excited state which happens in the thermodynamic limit. The latter is more common and has a closer analogy to classical phase transitions, while the former often occurs in conjunction with the latter. The QPT is usually accompanied by an abrupt change in the correlations in the ground state.

\subsection{\textbf{Example: the Mott-to-superfluid phase transition:}}
\label{subsubsec:bh}

\begin{figure}
\centering
\includegraphics[width=0.9\textwidth]{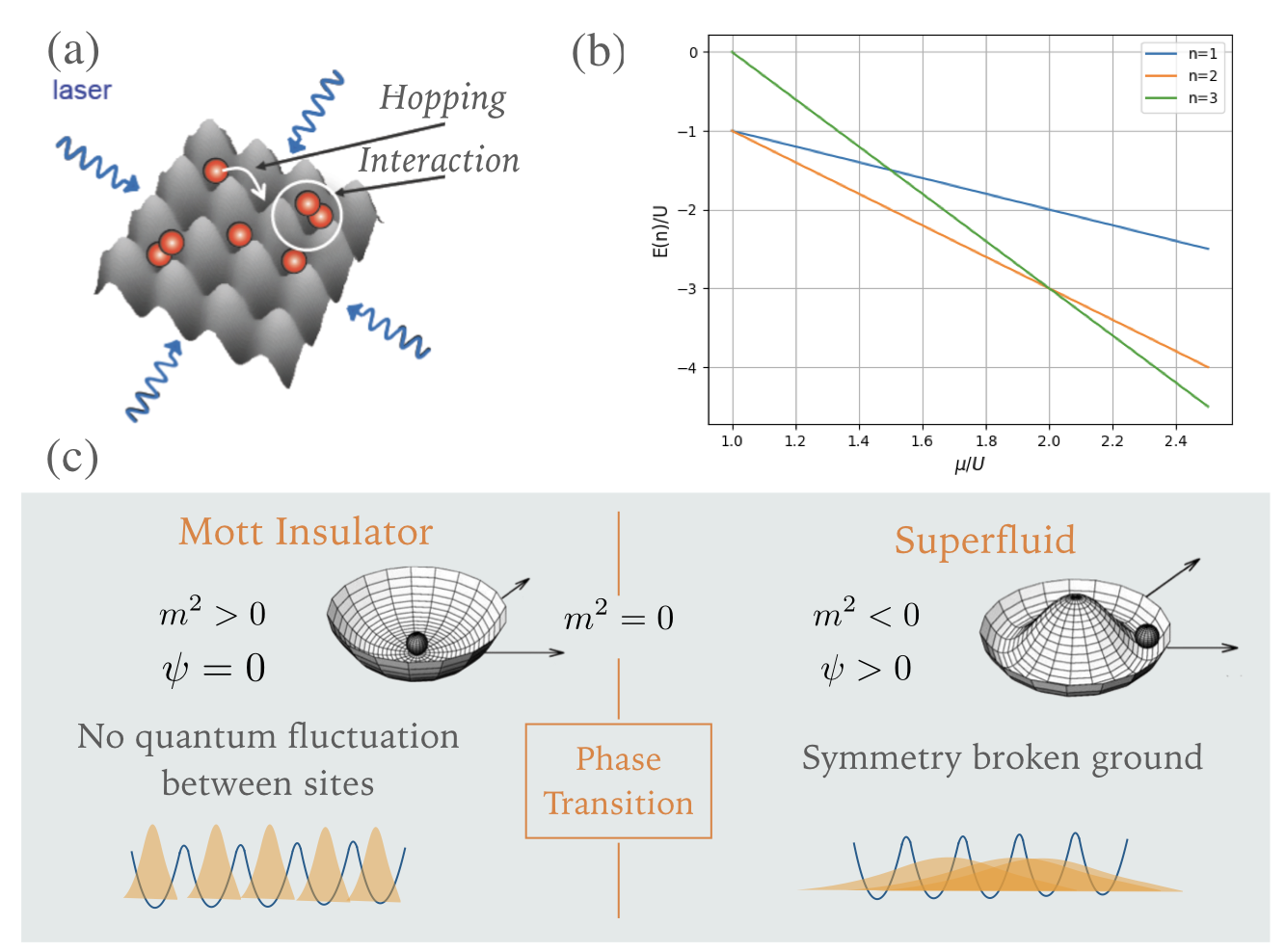}
\caption{\textbf{The Bose-Hubbard model.} \textbf{(a)} A sketch of cold atoms in optical lattices realizing the Bose-Hubbard model. \textbf{b} The mean-field energy as a function of $\mu/U$ for three different numbers of particles $n=1,2,3$. The mean-field energy landscape in the Mott and the superfluid phase are shown in \textbf{(b)} and \textbf{(c)}, respectively. }
\label{fig:bh_1}
\end{figure}

The Fermi-Hubbard model was originally proposed in 1963 by Hubbard \cite{Hubbard238} to approximately describe a conducting to an insulating QPT of electrons in solids. Its bosonic version was proposed in the same year by Gersch and Knollman \cite{PhysRev.129.959} and named the Bose-Hubbard model. The phase diagram of the latter was first calculated in 1989 \cite{PhysRevB.40.546} and the corresponding QPT was realized in cold atoms in optical lattices in \cite{2001_bloch_nat}. Specifically, the Bose-Hubbard (BH) model describes the system of $N$ bosonic particles moving on a lattice consisting of $L$ lattice sites,
\begin{equation}
\hat{H}_{\rm BH} = -J\sum_{\langle i,j\rangle}\hat{a}^\dagger_i \hat{a}_j-\mu\sum_{i=0}^{L-1}\hat{n}_j+\frac{U}{2}\sum_{i=0}^{L-1}\hat{n}_j(\hat{n}_j-1),
\label{eq:bose_hubbard_model}
\end{equation}
where $\hat{a}_j$ ($\hat{a}^\dagger_j$) is a bosonic annihilation (creation) operator at site $j$, $\hat{n}_j=\hat{a}^\dagger_j\hat{a}_j$ is the number operator at site $j$, $\langle..\rangle$ denotes the sum over nearest-neighbour, $J$ is the hopping strength between site $i$ and $j$, $\mu$ is the chemical potential, and $U$ is the on-site interaction. The first term in eq. (\ref{eq:bose_hubbard_model}) describes kinetic energy of particles, the second term determines the number of particles in the ground state, and the last term describes interaction between particles.

To understand different phases exhibited by the BH model, let us first consider the limit $J=0$.  In this case, $\hat{H}_{\rm BH}$ reduces to the sum of on-site Hamiltonians, \textit{i.e.}, $\hat{H}_{\rm BH}=\sum_j \hat{h}_j$, where $\hat{h}_j = -\mu \hat{n}_j+\frac{U}{2}\hat{n}_j(\hat{n}_j-1) $. Hence, the ground state takes the form of a product state $|G\rangle=\prod_j|n\rangle_j$, where $|n\rangle_j$ is an n-particle Fock state at site $j$, \textit{i.e.}, $\hat{n}_j|n\rangle_j=n|n\rangle_j$. The corresponding ground state energy is $E(n)=L\left[-\mu n +\frac{U}{2}n(n-1)\right]$. As shown in fig. (\ref{fig:bh_1}-b), there are  different level crossing between states with different integer fillings $n$ for $\mu = nU$. The ground state has an energy gap, hence they should be stable against small changes in the Hamiltonian such as small tunneling. This integer-filling ground states are known as Mott insulating states.

Another phase can be revealed by considering the limit $U=0$, where the Hamiltonian is reduced to the tight-binding model $\hat{H}_{\rm tight-binding}=-J\sum_{\langle i,j\rangle}\hat{a}^\dagger_i \hat{a}_j-\mu\sum_j\hat{n}_j$. The Hamiltonian can be diagonalized by applying the quantum Fourier transform $\hat{a}_k=\frac{1}{\sqrt{L}}\sum^{ikj}\hat{a}_j$. The Hamiltonian is then casted into the form $\hat{H}_{\rm tight-binding}=\sum_{k}\omega_k\hat{a}^\dagger_k\hat{a}_k$, where $\omega_k$ is a constant depending on $\mu$ and $k$. For example, for a one-dimensional periodic lattice, we have $\omega_k=-\mu-2J\cos(k)$. For $\mu, J>0$, the ground state is then $|G\rangle=(\hat{a}^\dagger_{k=0})^N|000...\rangle$ which is a product state in the momentum space, not in the position space as in the Mott. This state is known as the superfluid state.

\subsection{\textbf{The mean-field phase diagram:}}

Now let us consider an approximate method to calculate the many-body phases for the full range of $J/U$. The key idea is to decompose $\hat{a}_j=\psi+\delta\hat{a}_j$, where $\delta\hat{a}_j$ is the fluctuation from the mean value $\psi\equiv \langle\hat{a}_j\rangle$. The mean-field approximation proceeds by dropping the second order terms in $\delta\hat{a}_j$ in the Hamiltonian, assuming that correlations between sites can be ignored. This approximation becomes exact in infinite dimensions, but often fails in one-dimension. To see how this leads to the sum of approximated on-site Hamiltonians, let us write the hopping term as
\begin{align}
\hat{a}^\dagger_i\hat{a}_j + H.c. &= (\psi^*+\delta\hat{a}^\dagger_i)(\psi+\delta\hat{a}_j)+H.c. \nonumber \\
& = \psi^*\psi + \delta \hat{a}^\dagger_i \psi + \psi^*\delta \hat{a}_j+\delta\hat{a}_i\delta\hat{a}_j+H.c. \nonumber \\
& \approx \psi^*\psi +\delta\hat{a}^\dagger_i\psi+\psi^*\delta \hat{a}_j+H.c. \nonumber \\
& \approx \psi^*\psi + (\hat{a}^\dagger_i-\psi^*)\psi + (\hat{a}_j-\psi)\psi^*+H.c. \nonumber \\
& \approx \hat{a}^\dagger_i\psi +\hat{a}^\dagger_j\psi-\psi\psi^*+H.c..
\end{align}

\begin{figure}
\centering
\includegraphics[width=0.7\textwidth]{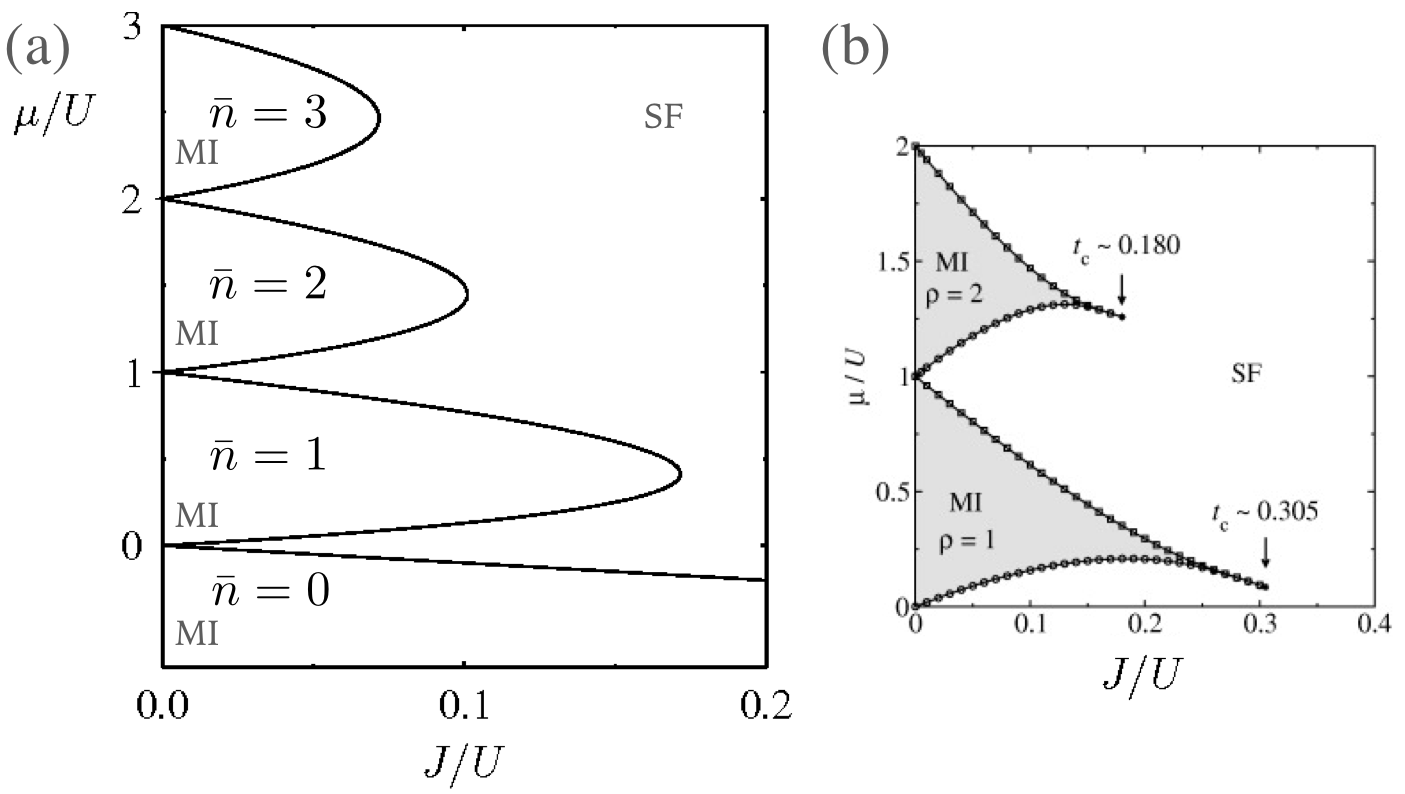}
\caption{\textbf{Phase diagram of the Bose Hubbard model}. The mean-field phase diagram showing the Mott and the superfluid phase is shown in \textbf{(a)}. A more exact phase diagram calculated from DRMG for the one dimensional system is shown in \textbf{(b)}. The result is reproduced from ref. \cite{0295-5075-93-3-30002}}
\label{fig:bh_2}
\end{figure}

The full Hamiltonian is then written as 
\begin{equation}
\hat{H}_{\rm BH}\approx\hat{H}_{\rm BH}^{\rm MF}=\sum_j (\hat{h}_j^{(0)}+\hat{V}_j)
\nonumber
\end{equation}
where $\hat{h}_j^{(0)} = -\mu \hat{n}_j +\frac{U}{2}\hat{n}_j(\hat{n}_j-1)-Jz\psi^*\psi $ and $\hat{V}_j  = -Jz(\psi^*\hat{a}_j+\psi\hat{a}^\dagger_j)$ with $z$ being the coordination number, or the number of sites connected to site $j$ via the hopping term. We then write the mean-field energy as 
\begin{equation}
E_n = E_n^{(0)}+E_n^{(1)}+E_n^{(2)}+... . 
\nonumber
\end{equation}
By doing perturbation theory with respect to the $V-$term, the zero order eigenenergy is
\begin{equation}
E^{(0)}_n = \left\{ \begin{array}{ll}
 0 &\mbox{ for $\mu<0$} \\
  -\mu n+\frac{U}{2}n(n-1)+Jz\psi^2 &\mbox{ for $U(n-1) <\mu <U n$}
       \end{array} \right.
       \nonumber
\end{equation}
The first order eigenenergy is zero $E^{(1)}_1=\langle n | \hat{V}_j | n \rangle=0$. The second order eigenenergy is 
\begin{equation}
E^{(2)}_n=\psi^2\sum_{n'}\frac{|\langle n |\hat{V}_j |n'\rangle|^2}{E^{(0)}_n-E^{(0)}_{n'}}=(Jz\psi)^2\left(\frac{u}{U(n-1)-\mu}+\frac{n+1}{\mu-Un}\right).
\nonumber
\end{equation}
The mean-field energy is then $E_n=\text{const.}+m^2\psi^2+...$ where
\begin{equation}
\frac{m^2}{Jz}=1+\frac{n}{\tilde{U}(n-1)-\tilde{\mu}}+\frac{n+1}{\tilde{\mu}-\tilde{U}n},
\nonumber
\end{equation}
with $\tilde{\mu}=\mu/Jz$ and $\tilde{U}=U/Jz$.

As shown in fig. (\ref{fig:bh_1}-c) and fig. (\ref{fig:bh_1}-d), for $m^2>0$ $E_n$ is minimized when $\psi=0$. Hence the ground state has $U(1)$ symmetry, \textit{i.e.}, invariant under the transformation $\psi\to e^{i\theta}\psi$, and no quantum fluctuation between sites corresponding to the Mott insulating states. For $m^2<0$, $E_n$ is minimized when $\psi\neq 0$, implying that the ground state has a broken $U(1)$ symmetry corresponding to the superfluid state. The phase boundary can be computed by solving the equation $m^2=0$. The corresponding mean-field phase diagram is shown in fig. (\ref{fig:bh_2}-a). Fig. (\ref{fig:bh_2}-b) shows the phase diagram for a one-dimension Bose-Hubbard lattice calculated by a more exact density-matrix-renormalization-group (DMRG) technique \cite{0295-5075-93-3-30002}, taking into account correlations between sites. We can see that the mean-field theory can give a qualitative approximation of the phase diagram.


\section{Quantum many-body phases of light}
\subsection{\textbf{Light-matter interaction:}}

Having discussed phases of matter, we now turn to experimental realization of photon-photon interactions and how many-body phases of light can emerge. Engineering strong interactions at progressively low light intensity has been one of the greatest challenges in optical science. In classical regime, photon-photon interaction is achieved by shining an intense light beam to a non-linear material so that optical properties of the material such as refraction and absorption are modified and, in turn, lead to power-dependent light propagation through the material \cite{2009_boyd}. Specifically, polarization $\textbf{P}$ of non-linear media, defined as dipole moments per unit volume, can be written as
\begin{equation}
\textbf{P}/\epsilon_0=\chi^{(1)}\textbf{E}+\chi^{(2)}\textbf{E}^2+\chi^{(3)}\textbf{E}^3+... ,
\label{eq:polarization}
\end{equation}
where $\epsilon_0$ is the electric permittivity of free space, $\chi^{i}$ is the electric susceptibility of order $i^{\rm th}$, and $\textbf{E}$ is the input electric field. The higher order terms account for non-linear optical phenomena such as second or higher harmonic generation, sum-frequency generation, self-focusing, and optical solitons. However, as the light intensity is weaker, the higher order terms in  eq. (\ref{eq:polarization}) are suppressed, and eventually, the material only exhibits a linear response, making it difficult to achieve strong interaction at a few photon levels. 

Another way to see this is to consider the probability $p$ of one photon getting absorbed by an atom. At resonance, this probability is maximized and proportional to the ratio between the wavelength of light squared ($\lambda^2$) and the transverse area of the laser beam ($d^2$), \textit{i.e.}, $p\sim \lambda^2/d^2$. The number of atoms required to modify one photon is then $N\approx 1/p$. Due to the diffraction limit that prevents the focusing of the light beam below the wavelength, this probability is typically small $p\ll 1$. Recent experiments have achieved $p\approx 0.01 - 0.1$ by concentrating laser light to a small area \cite{Darqui454,2007_sandoghdar_natphy, 2008_kurtsiefer_natphy,PhysRevLett.107.133002}.

In the limit $p\to 1$, the presence of one atom can substantially modify a single incident photon. Since a single two-level atom can only absorb one photon at a time, a pair of incident photons will experience an atomic response that is very different from that of a single photon, hence realizing nonlinearity at the two-photon level. One way to achieve this is to use a reflective cavity that enhances the interaction probability $p$ by the number of bounces, $F$, that the photon makes inside the cavity before leaking out. The probability $p$ approaches unity when $\eta\gg1$, where $\eta=F\lambda^2/d^2$ is the cooperativity.

\subsubsection{\textbf{Field quantization: mode of a simple optical resonator:}}
\label{app:field}

\begin{figure}
\centering
\includegraphics[height=3cm]{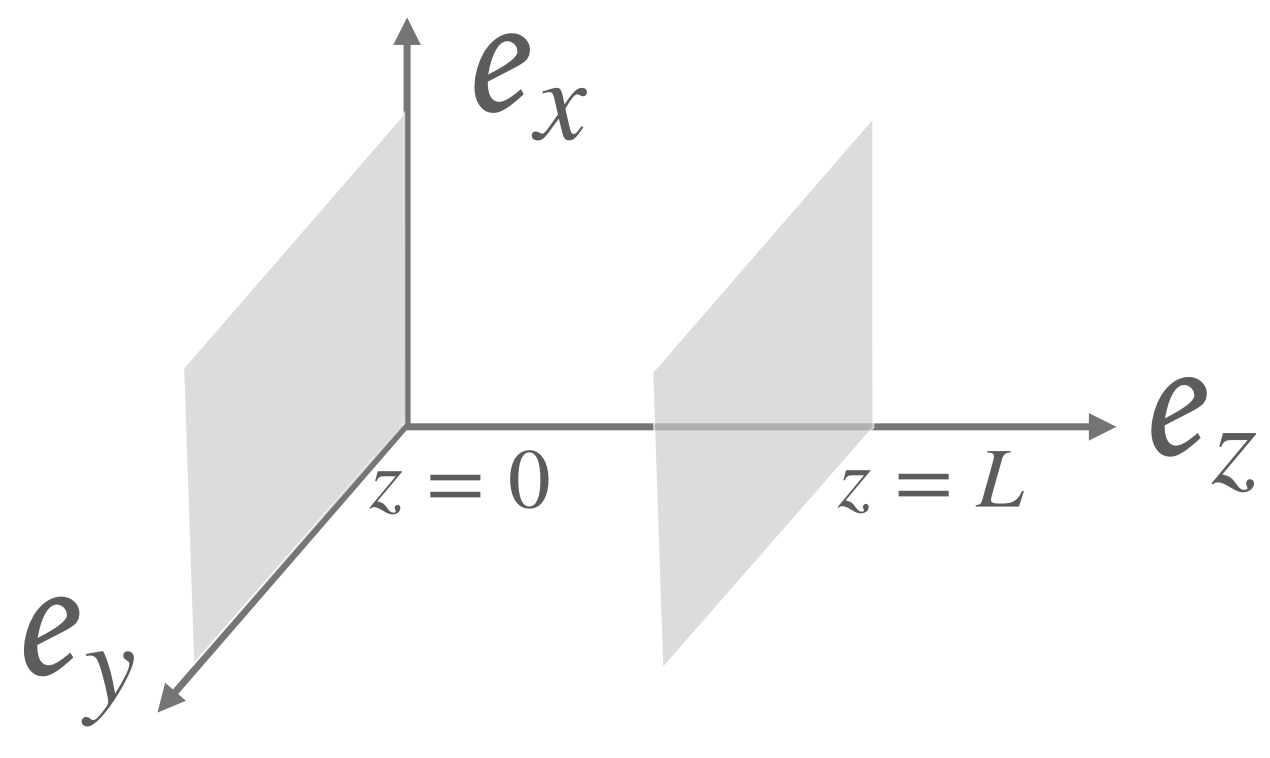}
\caption{A sketch of a simplified optical cavity consisting of two plane mirrors at $z=0$ and $z=L$.}
\label{fig:mirror}
\end{figure}

Let us consider an optical cavity consisting of two parallel perfectly reflecting mirrors, lying on the x-y plane at $z=0$ and $z=L$. The electric field and the magnetic field inside the cavity take the form
\begin{align}
\textbf{E}(\textbf{r},t) &= \textbf{e}_xE_x(z,t),  \nonumber\\
\textbf{B}(\textbf{r},t) &= \textbf{e}_yB_y(z,t),\nonumber
\end{align}
respectively. By solving Maxwell's equations assuming the boundary conditions $E_x(0,t)=E_x(L,t)=0$, we get
\begin{align}
E_x^{(m)}(z,t) &= E_0\sin(\omega^{(m)}_{c}t)\sin(k_mz), \nonumber\\
B_y^{(m)}(z,t) &=B_0\cos(\omega^{(m)}_{c}t)\cos(kz),\nonumber
\end{align}
where $k_m=m\pi/L$, $m$ is a positive integer, $\omega^{(m)}_{c}=ck_m$, $c$ is the speed of light, $E_0$ is the amplitude of the electric field, and $B_0=E_0/c$ is the amplitude of the magnetic field. The energy of the electromagnetic field inside the cavity is 
\begin{align}
E^{(m)} &= \frac{1}{2}\int dV\left[\epsilon_0|E^{(m)}_x(z,t)|^2+\frac{1}{\mu_0}|B^{(m)}_y(z,t)|^2\right] \nonumber \\
& = \frac{1}{4}\epsilon_0E^2_0V\left(\sin^2(\omega^{(m)}_{c}t)+\cos^2(\omega^{(m)}_{c}t)\right) \nonumber \\
& = \frac{1}{2}\left[(p^{(m)}(t))^2+(\omega^{(m)}_c)^2(q^{(m)}(t))^2\right],\nonumber
\end{align}
where $V$ is the volume of the cavity, $q^{(m)}(t)=\sqrt{\frac{\epsilon_0V}{2(\omega^{(m)}_c)^2}}E_0\sin(\omega^{(m)}_ct)$, and $p^{(m)}(t)=\sqrt{\frac{V}{2\mu_0}}B_0\cos(\omega^{(m)}_{c}t)$. One can see that the energy $E^{(m)}$ takes the form of the energy of a simple Harmonic oscillator where $p^{(m)}(t)$ and $q^{(m)(t)}$ are position and momentum coordinates, respectively. From now one we will consider the lowest mode $m=1$ and drop the superscript $(m)$.

Next, we perform second quantization by promoting $p(t)$ and $q(t)$ to operators, \textit{i.e.},
\begin{equation}
E^{(m)}\to \hat{H}_{\rm cavity} = \frac{1}{2}\left(\hat{p}^2+\omega^2_c\hat{q}^2\right),
\nonumber
\end{equation}
where $\left[\hat{q},\hat{p}\right]=i\hbar$. We then define the ladder operators as
\begin{align}
\hat{a}^\dagger = \frac{\omega_c \hat{q}-i\hat{p}}{\sqrt{2\hbar \omega_c}}, \nonumber \\
\hat{a} = \frac{\omega_c \hat{q}+i\hat{p}}{\sqrt{2\hbar \omega_c}}, \nonumber
\end{align}
where $\left[\hat{a},\hat{a}^\dagger\right]=1$. The Hamiltonian is then written as $\hat{H}_{\rm cavity} = \hbar \omega_c(\hat{a}^\dagger\hat{a}+\frac{1}{2})$ which is the form of a quantum harmonic oscillator. We will set $\hbar=1$ for now on for simplicity.

\subsubsection{\textbf{The Jaynes-Cummings interaction:}}
\label{sec:many_body_physics_with_interacting_photons}

\begin{figure}
\centering
\includegraphics[width=0.9\textwidth]{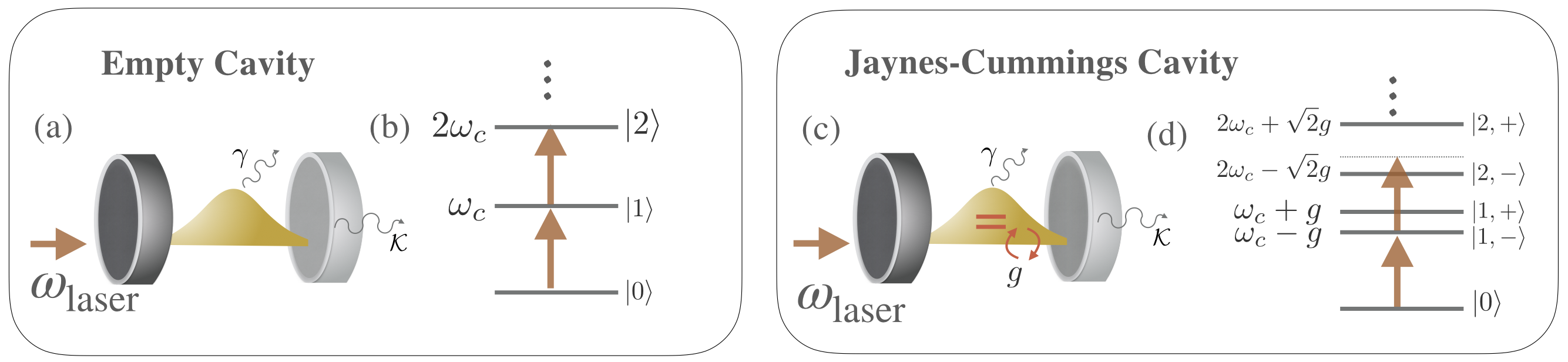}
\caption{\textbf{The Jaynes-Cummings model.} \textbf{(a)} A sketch of an empty cavity with \textbf{(b)} its linear spectrum. When the modes of the cavity coupled to the two-level system, the total system is described by the Jaynes-Cummings model \textbf{(c)}. \textbf{(d)} The resulting energy spectrum has non-linear splitting proportional to $g\sqrt{n}$. An external laser with frequency $\omega_{\rm laser}=\omega_c-g$ will be resonant with the one-excitation state $|1,-\rangle$ but off-resonant with the two-excitation state $|2,-\rangle$ leading to the photon blockade effect where only one photon can enter the cavity.}
\label{fig:cavity}
\end{figure}

The system consisting of a two-level atom interacting with photons trapped in an optical cavity, as shown in fig. (\ref{fig:cavity}), is described by the Hamiltonian 
\begin{equation}
\hat{H}_{\rm JC} = \hat{H}_{\rm cavity} +\hat{H}_{\rm atom} + \hat{H}_{\rm int},
\nonumber
\end{equation}
where
\begin{align}
\hat{H}_{\rm cavity} = \omega_c(\hat{a}^\dagger\hat{a}+\frac{1}{2})
\end{align}
is the Hamiltonian of the cavity. $\omega_c$ is the fundamental frequency of the cavity. The two-level atom is described as
\begin{equation}
\hat{H}_{\rm atom}=\omega_a\hat{\sigma}^+\hat{\sigma}^-,
\nonumber
\end{equation}
where $\hat{\sigma}^+=|e\rangle\langle g|$ and $\omega_a$ is the energy different between the two eigenstates. The atom interacts with the cavity mode by a dipole transition operator which is defined as $\hat{\textbf{d}}=\textbf{d}^*\hat{\sigma}^++\textbf{d}\hat{\sigma}^-$, where $\textbf{d}$ is the dipole moment. The interaction between the atom and the cavity is described by the dipole interaction.
\begin{align}
\hat{H}_{\rm int} &= -\hat{\textbf{d}}\cdot\hat{\textbf{E}}(z,t) \nonumber \\
&=-E_0(\hat{a}+\hat{a}^\dagger)\sin\left(\frac{\pi z}{L}\right)\hat{d} \nonumber \\
&= g(\hat{\sigma}^++\hat{\sigma}^-)(\hat{a}+\hat{a}^\dagger) \nonumber \\ 
&= g(\hat{\sigma}^\dagger\hat{a}+\hat{\sigma}^-\hat{a}+\hat{\sigma}^+\hat{a}^\dagger+\hat{\sigma}^-\hat{a}^\dagger), \nonumber
\end{align}
where $g=-E_0\sin(\frac{\pi z}{L})$ is the light-matter coupling constant. $E_0$ is the amplitude of the field in the cavity of length $L$. $z$ is the position in the cavity. We can see that $\hat{H}_{\rm int}$ contains terms that do not conserve the number of excitation. To see the effect of these terms, we move from the Schr\"odinger  picture into the interaction picture defined by $\hat{H}_{\rm cavity}+\hat{H}_{\rm atom}$, \textit{i.e.},
\begin{align}
\hat{H}_{\rm int}(t) = &e^{i(\hat{H}_{\rm cavity}+\hat{H}_{\rm atom})t}\hat{H}_{\rm int}e^{-i(\hat{H}_{\rm cavity}+\hat{H}_{\rm atom})t} \nonumber \\\
= &g\left(\hat{a}\hat{\sigma}^-e^{-i(\omega_c+\omega_a)t}+\hat{a}\hat{\sigma}^+e^{i(\omega_a-\omega_c)t} + H.c. \right).
\end{align}
The terms $\hat{a}^\dagger \hat{\sigma}^-$ and $\hat{a}\hat{\sigma}^+$ describe an emission and an an absorption process, respectively. They oscillate with a slow frequency $\omega_a-\omega_c$, while the counter-rotating terms $(\hat{a}\hat{\sigma}^-e^{-i(\omega_c+\omega_a)t}+H.c.)$ do not conserve number of excitations and quickly oscillates. When $|\omega_c-\omega_a|, g\ll \omega_c+\omega_a$, the latter terms can be ignored, giving rise to a solvable model known as the Jaynes-Cummings (JC) model  \cite{2008_milburn_springer,1993_knight_jmo} first envisioned in 1963. This approximation is known as the rotating-wave approximation. The JC model is then written as
\begin{equation}
\hat{H}_{\rm JC}=\omega_a \hat{\sigma}^+ \hat{\sigma}^- + \omega_c \hat{a}^\dagger \hat{a} + g (\hat{a}^\dagger \hat{\sigma}^- + \hat{a}\hat{\sigma}^+).
\nonumber
\end{equation}

\subsubsection{\textbf{Eigenstates of the Jaynes-Cummings model:}}

To obtain the eigenstates and the eigenenergies of the JC model, we first notice that the Hamiltonian $\hat{H}_{\rm JC}$ commutes with the total number excitation operator 
\begin{equation}
\hat{N}=\hat{\sigma}^+\hat{\sigma}^-+\hat{a}^\dagger\hat{a}. 
\nonumber
\end{equation}
For $n$ excitations, there are only two possible states which are (i) $|\psi_1\rangle = |n-1,e\rangle $ with $n-1$ photons in the cavity and the atom is in the excited state and (ii) $|\psi_2\rangle = |n,g\rangle$ with $n$ photons in the cavity and the atom is in the ground state. The matrix elements of $\hat{H}_{\rm JC}$ in this subspace, 
\begin{equation}
H_{\rm JC, ij}^{(n)}= \langle\psi_i |\hat{H}_{\rm JC}  |\psi_j\rangle
\label{eq:rabi}
\end{equation}
for $i,j\in\{1,2\}$, are written as
\begin{equation}
\hat{H}_{\rm JC}^{(n)}=\begin{pmatrix}(n-1)\omega_c +\omega_a& g\sqrt{n} \\ g\sqrt{n} & n\omega_c\end{pmatrix}.
\nonumber
\end{equation}
Diagonalizing this Hamiltonian, we obtain the energy eigenstates as
\begin{equation}
E_{\pm}(n) = \omega_c \left(n-\frac{1}{2}\right)\pm \frac{1}{2}\sqrt{(\omega_a-\omega_c)^2+g^2n}.
\label{eq:jcenergy}
\end{equation}
with the energy eigenstates
\begin{align}
|n,+\rangle &= \cos\left(\frac{\alpha_n}{2}\right)|n-1,e\rangle+\sin\left(\frac{\alpha_n}{2}\right)|n,g\rangle, \nonumber \\
|n,-\rangle &= -\sin\left(\frac{\alpha_n}{2}\right)|n-1,e\rangle+\cos\left(\frac{\alpha_n}{2}\right)|n,g\rangle \nonumber,
\end{align}
where $\alpha_n=\tan^{-1}\left(2g\sqrt{n}/(\omega_a-\omega_c)\right)$. Excitations in $|n,\pm\rangle$ are a collective mode of photonic and atomic excitations called a polariton. The cavity is said to be a non-linear cavity because its eigenenergies now have non-linear dependence in $n$. The non-linearity becomes maximized at resonance $\omega_a=\omega_c$, \textit{i.e.} ,$E_{\pm}(n)= \omega_c (n-1/2) \pm \frac{1}{2}g\sqrt{n+1}$.  In the large detuning limit $|\omega_a-\omega_c|\gg g\sqrt{n}$, the eigenenergies becomes approximately linear in $n$, \textit{i.e.}, 
\begin{equation}
E_{\pm}(n)\approx \omega_c n \pm \frac{1}{2}|\omega_a-\omega_c|.
\nonumber
\end{equation}
In this limit, the cavity modes are decoupled from the atom. In addition, the spectrum is also approximately linear for a large number of photons $n\gg 1$ where $\sqrt{n+1}\approx\sqrt{n}$ since an energy gap between adjacent energy levels are approximately the same, e.g. 
\begin{equation}
E_{\pm}(n+2)-E_{\pm}(n+1)\approx E_{\pm}(n+1)-E_{\pm}(n)\approx \omega_c\pm \frac{1}{2}g\sqrt{n}.
\nonumber
\end{equation} 

\subsubsection{\textbf{Early experimental realizations of strong light-matter coupling:}}

\begin{figure}
\centering
\includegraphics[width=0.9\textwidth]{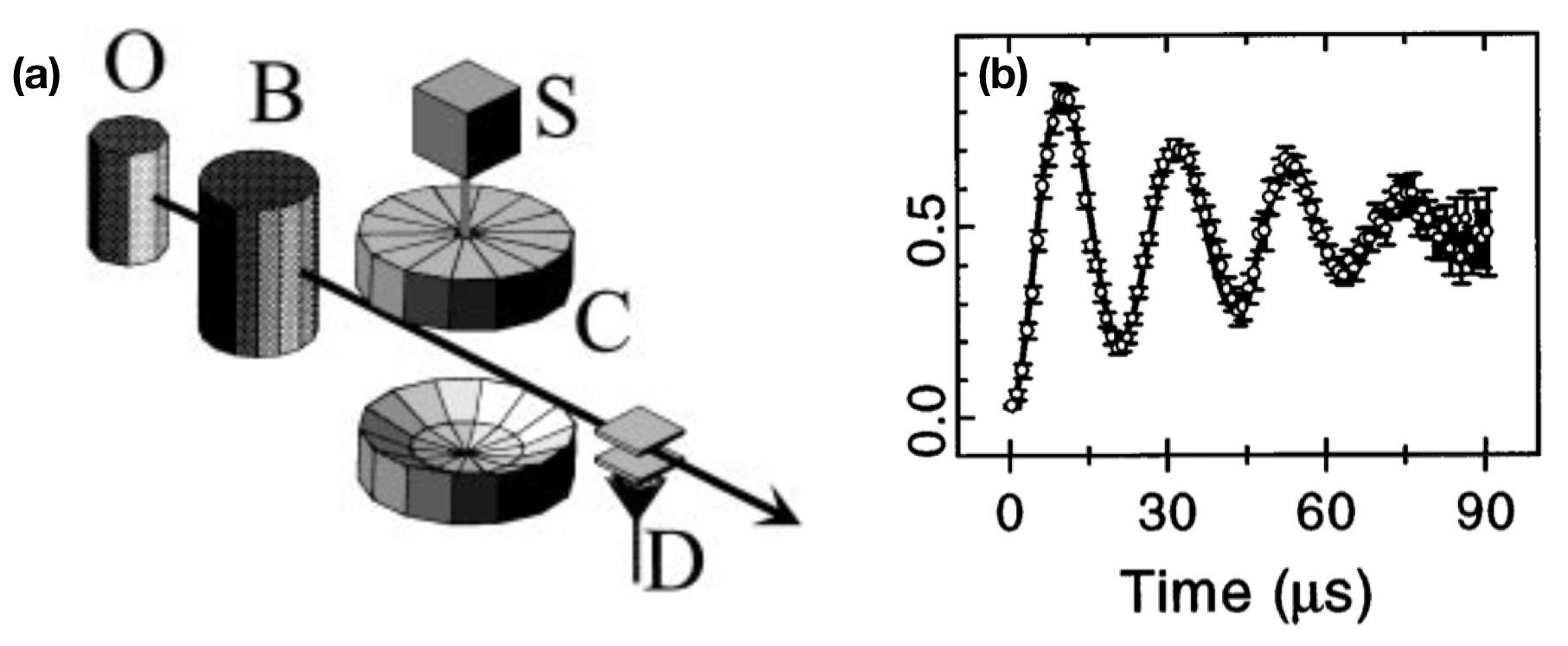}
\caption{\textbf{Quantum Rabi oscillation in a microwave cavity.} \textbf{(a)} A simple diagram of the experiment. Rubidium atoms effuse from the oven O and circular Rydberg atoms are prepared in the box B. The atoms cross the cavity C made of two superconducting mirrors.  \textbf{(b)} The observed Rabi oscillation according to the Hamiltonian in eq. (\ref{eq:rabi}). The result is reproduced from ref. \cite{PhysRevLett.76.1800}.}
\label{fig:rabi}
\end{figure}

Signatures of the atom-cavity interaction were first observed in the 1980s via the change in the spontaneous emission rate of an atom when placed in a cavity \cite{PhysRevLett.50.1903,PhysRevLett.55.2137,PhysRevLett.58.666}. The cavity mode is said to be `strongly coupled' to the atom when the light-matter coupling is much larger than dissipation rate both to the input and the output waveguides $\mathcal{K}$ and to free space $\gamma$, \textit{i.e.}, 
\begin{equation}
g^2>\mathcal{K}\gamma.
\nonumber
\end{equation}
In this limit, a single photon in the cavity has enough coherent time to allow reversible exchange between the atomic and the photonic excitation before irreversibly leaking out the cavity. Signatures of the strong light-matter interaction were first observed in the optical regime in 1992 via normal mode splitting \cite{PhysRevLett.68.1132} and in the microwave regime in 1996 via quantum Rabi oscillation \cite{PhysRevLett.76.1800}, see fig. (\ref{fig:rabi}). The former led to the first experimental demonstration that single atoms can introduce a phase shift to a single photon by approximately $\pi/10$ \cite{PhysRevLett.75.4710}, while the latter led to the generation of Einstein-Podolsky-Rosen pair of atoms in a controllable manner \cite{PhysRevLett.79.1}. In superconducting systems, strong coupling between a single artificial atom and a single microwave photon was later realized in 2004 \cite{2004_wallraff_nat}.

\subsubsection{\textbf{Photon blockade effect:}}

\begin{figure}
\centering
\includegraphics[width=0.9\textwidth]{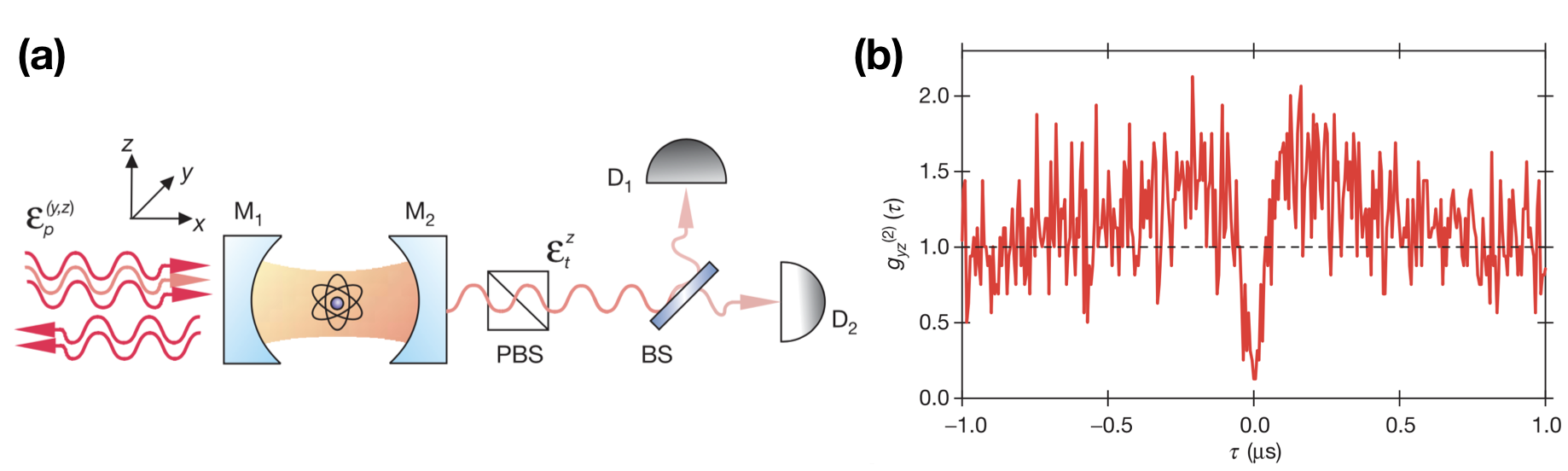}
\caption{\textbf{Experimental realization of photon blockade.} \textbf{(a)} A simple diagram of the experiment. \textbf{(b)} The intensity correlation function $g^{(2)}(\tau)$ as a function of the time delay $\tau$ between two photons. $g^{(2)}(\tau)$ drops to near zero at $\tau=0$, indicating the probability of detecting two photons at the same time is strongly suppressed. The result is reproduced from ref. \cite{2005_kimble_nat}.}
\label{fig:g2}
\end{figure}

Photon blockade refers to a situation in which interaction between photons is so strong that the presence of a single photon in a cavity can completely `block' another photon from entering the cavity. The term is used in analogy to the Coulomb blockade effect \cite{Averin1986} where a single electron on a small metallic or semiconductor device can block the flow of another electron when the charging energy is much larger than the thermal energy. To understand photon blockade in the Jaynes-Cummings model, imagine the cavity and the atom is initially in the vacuum state and the ground state, respectively, \textit{i.e.}, $|0,g\rangle$. A laser beam is then shined to the system with the frequency that is resonant with one of the one-excitation eigenstates, e.g., 
\begin{equation}
\omega_{\rm laser}=E_-(1)=\omega_c-\frac{1}{2}g
\nonumber
\end{equation}
(assuming $\omega_a=\omega_c$). Due to the resonance condition, the first photon that enters the cavity will excite the cavity to the state $|1,-\rangle$. However, the frequency of the second photon is now off-resonant with that of the two-excitation state $|2,-\rangle$,
\begin{equation}
E_\pm(2)-E_-(1)=\omega_c\mp\frac{1}{2}g(\sqrt{2}\mp1)\neq\omega_{\rm laser}. 
\nonumber
\end{equation}
Hence the second photon is prevented from entering the cavity and the two photons effectively `repel' each other.  The first experimental breakthrough that showed direct signatures of photon blockade was done in 2005 \cite{2005_kimble_nat}, through the anti-bunching statistics of transmitted photons, see fig. (\ref{fig:g2}). The result marks an exciting new era in nonlinear quantum optics. 

\subsubsection{\textbf{Quantum nonlinear optics with atomic ensembles:}}

Before moving on to quantum many-body body physics with light, we would like to mention an alternative approach to engineer strong light-matter interaction where photons are stored in an ensemble of atoms which exhibits the so-called electromagnetically induced transparency (EIT) \cite{Schmidt:96}. In EIT, the optical response of an otherwise opaque atomic gas is modified by an extra control field. This control field is strong and induces coherent coupling between a weak probe pulse and atomic states leading to collective light-matter excitations, called polaritons. The latter results in a drastically reduced group velocity of the probe field, much reduced linear susceptibility $\chi^{(1)}$, and greatly enhanced nonlinear susceptibility $\chi^{(3)}$. Effective strong polariton-polariton interaction is then induced by exciting the atoms to the metastable Rydberg state with a high principal quantum number of approximately $100$ \cite{2012_vladan_nat}. The strong interaction between two Rydberg atoms that are separated by less than the blockade radius introduces an energy shift when two of them are excited. This energy shift is maximized when both controlled field and probe field are resonant with corresponding atomic states. The latter prevents both of them to get excited simultaneously. Effectively, each Rydberg atom behaves like a `superatom' consisting of $N_a$ atoms within the Rydberg radius but only one excitation, resulting in enhanced cooperativity of $\eta=N_a\lambda^2/d^2$.

In addition to the above Rydberg blockade, when the control laser is detuned from resonance, it is possible to use EIT to engineer effective distance-dependent interaction between photons. The attractive interaction between photons has been realized in this way and two-photon bound states have been observed \cite{vladan}. In the case of repulsive interaction, it is predicted that photon crystallization could be formed \cite{2014_lukin_natphy}.


\subsection{\textbf{Mott-to-superfluid transition of light in coupled resonator arrays: }}

\begin{figure}
\centering
\includegraphics[width=1.0\textwidth]{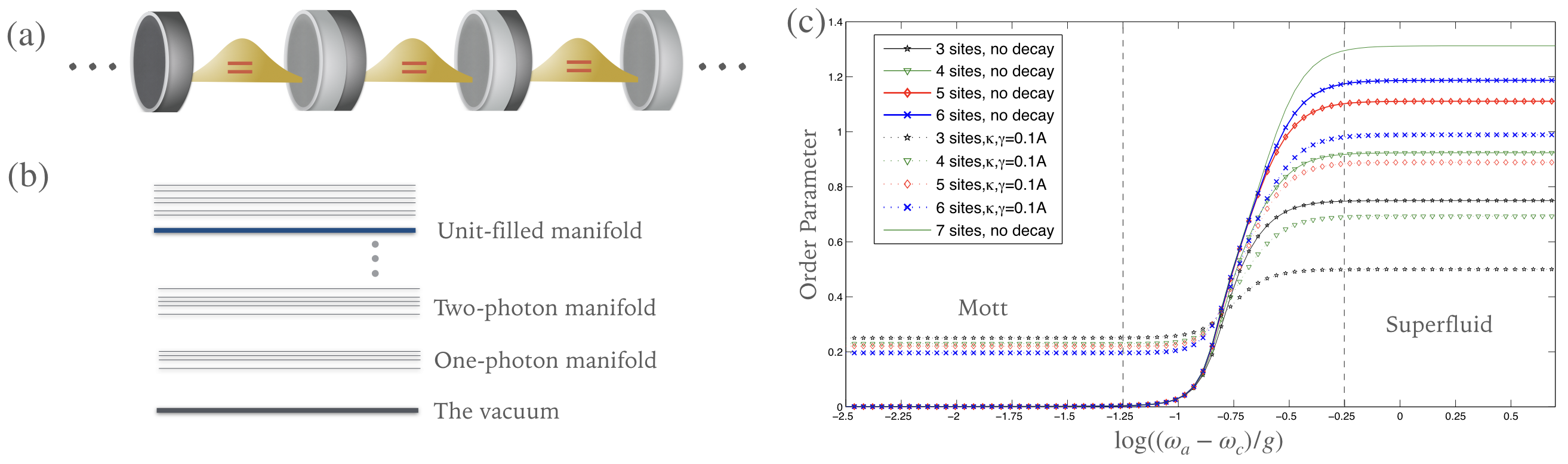}
\caption{\textbf{The Jaynes-Cummings-Hubbard model} \textbf{(a)} A sketch of a coupled cavity array, implementing the Jaynes-Cummings-Hubbard model. \textbf{(b)} Energy spectrum of the JCH model. \textbf{(c)} The order parameter $\text{Var}({N}_i)$ of the lowest-energy state in the unit-filled manifold as a function of detuning $(\omega_a-\omega_c)/g$ for 3-7 sites with and without decay. The order parameter exhibits a jump from zero to a finite value, corresponding to the Mott and the superfluid phase, respectively. The transition gets shaper with as the system's size is increased as expected from quantum phase transition. The results are reproduced from ref. \cite{2007_dimitris_pra}.}
\label{fig:mott}
\end{figure}

Having realized strong photon-photon interactions, it is natural to ask if photons can form many-body states in analogy to real atoms in solid state. Pioneer works explored this question by envisioning an array of coupled nonlinear cavities both with the Kerr type \cite{2006_hartmann_natphy} and the Jaynes-Cummings type \cite{2007_dimitris_pra, 2006_greentree_natphy}. The latter, as shown in fig. (\ref{fig:mott}-a), is described by the Hamiltonian, 
\begin{equation}
\hat{H}_{\rm JCH} =   \sum_{j=0}^{L-1}\left( \omega_a\hat{\sigma}^+_j\hat{\sigma}^-_j + \omega_c \hat{a}^\dagger_j \hat{a}_j + g (\hat{a}^\dagger_j \hat{\sigma}^-_j + \hat{a}_j\hat{\sigma}^+_j)\right) - J\sum_{j=0}^{L-2}\left(\hat{a}^\dagger_{j} \hat{a}_{j+1}+H.c.\right),
\nonumber
\end{equation}
where $L$ is the size of the system, $J$ is the hopping strength of photons between two adjacent cavities, $\sigma^+_j$ ($\sigma^-_j$) is the raising (lowering)operator for the atom at site $j$, and $\hat{a}_j$ ($\hat{a}^\dagger_j$) is a bosonic annihilation (creation) operator at site $j$. The model is known as the Jaynes-Cummings-Hubbard (JCH) model. $\hat{H}_{\rm JCH}$ commutes with the total number of excitations $\hat{N}=\sum_j\hat{N}_j$, where 
\begin{equation}
\hat{N}_j=\hat{a}^\dagger_j\hat{a}_j+\hat{\sigma}^+_j\hat{\sigma}^-_j. 
\nonumber
\end{equation}
Hence, as shown in fig. (\ref{fig:mott}-b), the eigenspectrum of $\hat{H}_{\rm JCH}$ are grouped into manifold labeled by the filling factor $\bar{n}=\langle \hat{N} \rangle/L$ where $\langle ...\rangle$ denotes an expectation value over a given eigenstate. It is important to recall that when deriving the Jaynes-Cummings interaction we have assumed that $\omega_a$ and $\omega_c$ are the largest energy scale in the system. This implies that the ground state of $\hat{H}_{\rm JCH}$ is the vacuum, as depicted in fig. (\ref{fig:mott}-b).

To observe many-body characteristics of $\hat{H}_{\rm JCH}$, one can consider the lowest energy state $|G\rangle_{\bar{n}=1}$ in the unit-filled manifold. At resonance and $g\gg J$, photon blockade prevents two photonic excitations at the same cavity, effectively switching off the hopping process and leading to the Mott-like ground state, \textit{i.e.}, 
\begin{equation}
|G\rangle_{\bar{n}=1}=|1,-\rangle\otimes|1,-\rangle...\otimes|1,-\rangle. 
\nonumber
\end{equation}
This state can be prepared by sending a $\pi/2$ pulse at frequency $\omega_a-g$ to each cavity to excite the vacuum state $|0,g\rangle$ to the lower polariton state $|-1\rangle$. To observe the superfluid behavior of photonic excitations, one can adiabatically switch on either the coupling $g$ or the detuning $\omega_a-\omega_c$ via, say, a Stark shift from an external field. In this limit, photon blockade is suppressed and the system is effectively described by the tight-binding model 
\begin{equation}
\hat{H}_{\rm JCH} \approx - J\sum_j\left(\hat{a}^\dagger_{j} \hat{a}_{j+1}+H.c.\right). 
\nonumber
\end{equation}
The Mott-to-superfluid phase transition can then be probed by measuring the fluctuation of the number of excitations, \textit{i.e.}, 
\begin{equation}
\text{Var}({N}_i)=\sqrt{\langle\hat{N}_j^2\rangle-\langle \hat{N}_j\rangle^2}, 
\nonumber
\end{equation}
see fig. (\ref{fig:mott}-c) . Note that, unlike atoms in the Bose-Hubbard lattice discussed in Sec. (\ref{subsubsec:bh}), the number of excitations, in this case, is conserved. Hence the order parameter $\langle a_i\rangle$ always vanishes both in the Mott and the superfluid phase.

\subsubsection{\textbf{The mean-field phase diagram:}}

\begin{figure}
\centering
\includegraphics[width=0.9\textwidth]{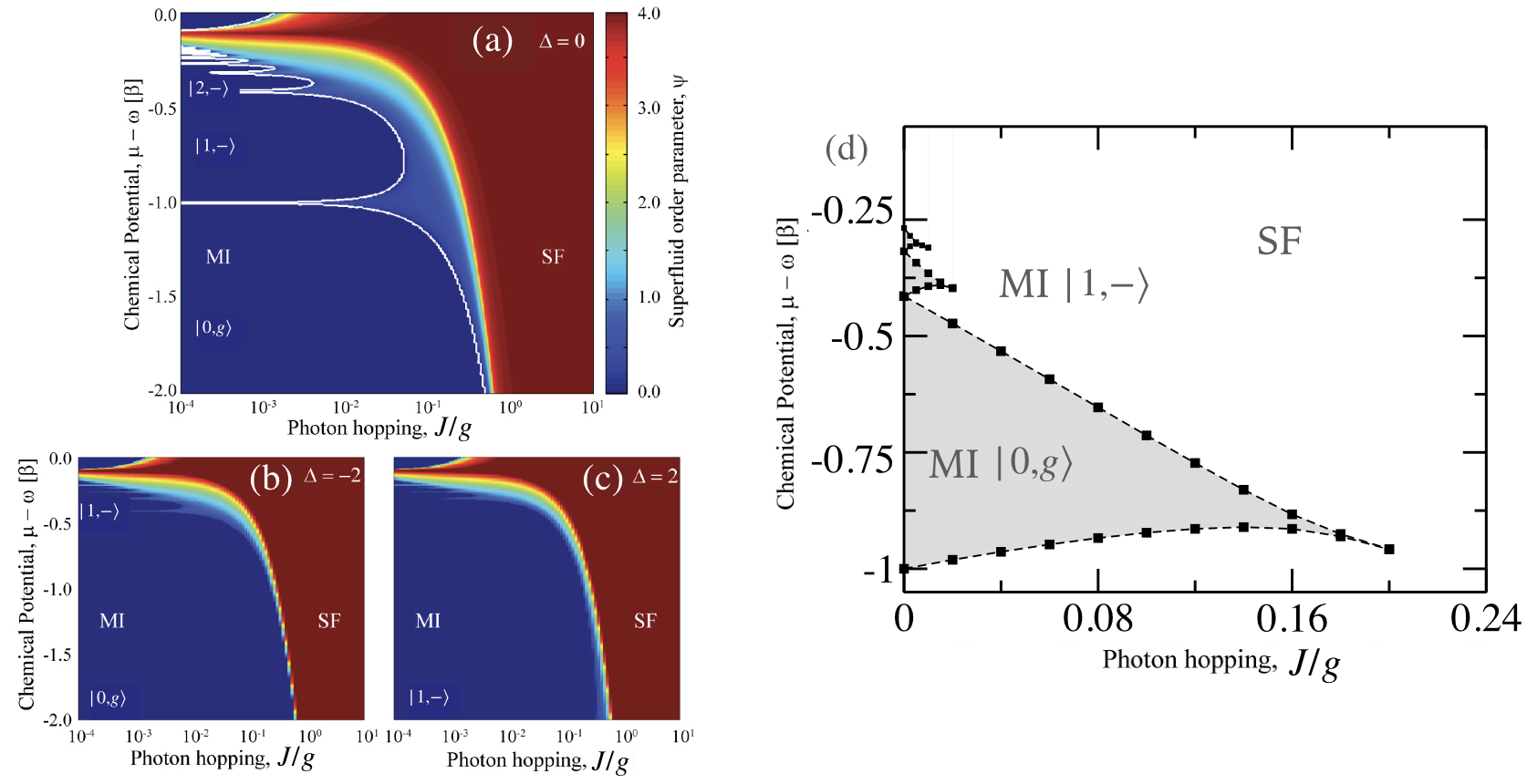}
\caption{\textbf{Phase diagram of the Jaynes-Cummings-Hubbard model}. The mean-field phase diagrams with different detuning $\Delta\equiv \omega_a-\omega_c=0,-2J, 2J$ are shown in \textbf{(a)-(c)}, respectively. The results are reproduced from ref. \cite{2006_greentree_natphy}. The DRMG phase diagram for the one-dimensional system with $\Delta=0$ is shown in (d). The result is reproduced from ref. \cite{PhysRevLett.99.186401}.}
\label{fig:jch_pd}
\end{figure}

To make a more explicit analogy to the atoms in the Bose-Hubbard lattice, one can introduce the chemical potential term to the JCH model \cite{2006_greentree_natphy}. Note that this is done \textit{by hand} since photons do not naturally have a chemical potential. Nevertheless, a possibility to engineer one has been proposed \cite{PhysRevB.92.174305}. The JCH model with the chemical potential is written as
\begin{equation}
\hat{H}_{\rm JCH}^{\rm GC} =  \hat{H}_{\rm JCH}-\mu\hat{N},
\nonumber
\end{equation}
where $\mu$ is the chemical potential and the label `GC' stands for the grand canonical potential. $\hat{H}_{\rm JCH}^{\rm GC}$ still commutes with $\hat{N}$. However, it is now possible that the ground state of the system is not the vacuum because the chemical potential term introduces an energy shift of $-\mu\bar{n}L$ to the excited states of $\hat{H}_{\rm JCH}$. This removes the need of restricting ourselves to the $\bar{n}=1$ manifold as before. 

We then proceed to calculate the ground-state phase diagram of $\hat{H}_{\rm JCH}^{\rm GC}$ by applying the mean-field approximation $\hat{H}_{\rm JCH}^{\rm GC}\approx\sum_j \hat{H}^{\rm MF}_{\rm JCH}(j)$, where 
\begin{align}
 \hat{H}^{\rm MF}_{\rm JCH}(j) &=  (\omega_c-\mu)\hat{a}^\dagger_j\hat{a}_j+(\omega_a-\mu)\hat{\sigma}^+\hat{\sigma}^-+g(\hat{a}^\dagger\hat{\sigma}^-+\hat{a}\hat{\sigma}^+) \nonumber\\ &-2J(\psi\hat{a}^2+\psi^*\hat{a}-|\psi|^2)
\end{align}
To numerically compute the phase diagram, we first write down $\hat{H}^{\rm MF}_{\rm JCH}(j)$ in a matrix form, keeping up to $n_{\rm max}$ excitations. For example, for $n_{\rm max}=1$ the matrix takes the form
\begin{equation}
 \hat{H}^{\rm MF}_{\rm JCH}(j) =\begin{pmatrix}
0 & 0 & -2J\psi \\
0 & \omega_a-\mu & g \\
-2J\psi^* & g & \omega_c-\mu 
\end{pmatrix},  
\nonumber
\end{equation}
where $|g,0\rangle=(1,0,0)$, $|e,0\rangle=(0,1,0)$, and $|e,1\rangle=(0,0,1)$. For $n_{\rm max}=2$ the matrix takes the form
\begin{equation}
 \hat{H}^{\rm MF}_{\rm JCH}(j) =\begin{pmatrix}
0 & 0 & -2J\psi & 0&0\\
0 & \omega_a-\mu & g &-2J\psi & 0\\
-2J\psi^* & g & \omega_c-\mu  & 0 & -2\sqrt{2}J\psi \\
0 & -2J\psi & 0 & \omega_a+\omega_c-2\mu & \sqrt{2} g \\
0 & 0 & -2J\psi & \sqrt{2}g & 2\omega_c-\mu
\end{pmatrix}.  
\nonumber
\end{equation}
The next step is to numerically obtain the ground state energy as a function of the mean-field energy $E\left[\psi\right]$ and find $\psi_c$ that minimizes $E[\psi]$. The process is then repeated until $\psi_c$ is converged with $n_{\rm max}$. The mean-field phase diagram of $\hat{H}_{\rm JCH}^{\rm GC}$ as calculated in ref. \cite{2006_greentree_natphy} is shown in fig. (\ref{fig:jch_pd}-a). A more accurate phase diagram was calculated numerically using DMRG in ref. \cite{PhysRevLett.99.186401} and analytically in ref \cite{PhysRevA.80.053821}. 

\subsubsection{\textbf{Existing works on equilibrium many-body phases of interacting photons:}}

Following the pioneer works, there has been several work investigating various aspects of the JCH model including many-body dynamics \cite{PhysRevA.80.043842,PhysRevA.80.033612,0953-4075-44-11-115505}, ground-state entanglement \cite{PhysRevA.77.022103, PhysRevA.80.043825}, critical exponents at the phase transition \cite{PhysRevA.84.041608} and its applications for quantum information processing \cite{doi:10.1080/09500340701515120, PhysRevA.76.013819}. Phase transitions in the JCH model and the BH model have been shown to be in the same universality class. As shown in \cite{2007_dimitris_pra} in the Mott regime, the JCH model also simulates the standard XY spin model where the presence and the absence of a polariton in each cavity represent the state of spin up and down, respectively. Subsequent works also show that the anisotropic Heisenberg spin model can be simulated using coupled cavity arrays where each cavity contains a three-level system \cite{2007_hartmann_prl,2008_angelakis_el, 2009_li_ep,2012_sarkar_pb}. Topological pumping of interacting photons to reliably transport Fock states is discussed in ref. \cite{2016_tangpanitanon_prl}. Artificial gauge field for photons can be engineered in a 2D array using an external drive that controls the hopping phase of photons. When combined with photon blockade, the ground state of the system can be mapped to the Laughlin state, simulating the fractional quantum Hall state of light \cite{2008_angelakis_prl, 2011_girvin_njp}. Such combination has been realized with two microwave photons in a three-site superconducting circuit chip \cite{2016_roushan_natphy}.  

\subsubsection{\textbf{State-of-the-art experiments:}} Here we briefly review two experimental works that demonstrate exceptional control and a long coherent time of interacting photons in superconducting circuits.

\begin{figure}
\centering
  \includegraphics[width=0.9\textwidth]{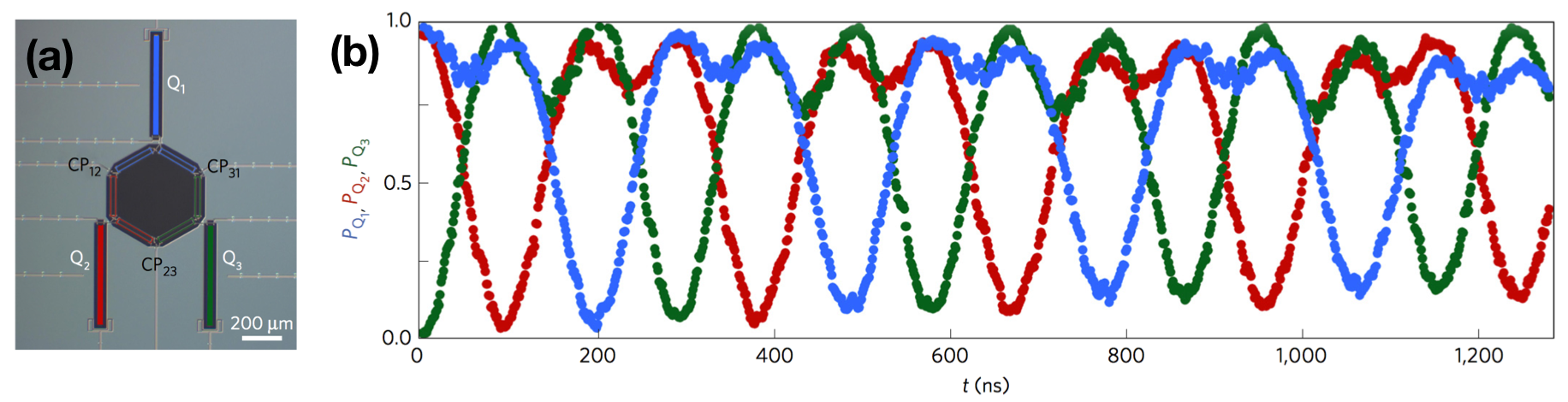}
\caption{\textbf{Chiral edge states of interacting photons in superconducting circuits.} \textbf{(a)} An optical image of the superconducting circuit made by standard nano-fabrication techniques. It can be described by the Bose-Hubbard Hamiltonian as shown in eq. (\ref{eq:chiral}). \textbf{(b)} Time evolution showing chiral current of two interacting photons.  $P_{Q_j}$ is the probability of finding one photon at site $j$. Two photons are initialized at sites $1$ and $2$ by applying a $\pi/2$ pulse to the corresponding sites. The chiral current can be understood by looking at the current of the hole, initially located at site 3. The results are reproduced from ref. \cite{2016_roushan_natphy}. }
\label{fig:chiral}
\end{figure}

\textit{Chiral ground-state currents of interacting photons in a synthetic magnetic field \cite{2016_roushan_natphy}.} Using two interacting photons in a three-site superconducting circuit (see fig. (\ref{fig:chiral}-a)), the authors have realized two basic ingredients for simulating fractional quantum Hall states of light, \textit{i.e.}, large interactions in the presence of a large magnetic field. The resulting ground states exhibit chiral current of two hardcore photons, where the probability of two photons being in the same site is strongly suppressed, see fig. (\ref{fig:chiral}-b). The artificial magnetic field is created by periodically modulating hopping strength between two sites \cite{PhysRevA.87.062336}, \textit{i.e.},
\begin{equation}
\hat{H}^{\rm chiral}(t)=\sum_{j=1}^3\omega_j \hat{n}_j +\sum_{j,k}J_{jk}(t)(\hat{a}^\dagger_j\hat{a}_k+H.c.)-\frac{U}{2}\sum_{j=1}^3\hat{n}_j(\hat{n}_j-1),
\label{eq:chiral}
\end{equation}
where 
\begin{equation}
J_{jk}(t) = J_0 \cos(\Delta_{jk}+\phi_{jk}),
\nonumber
\end{equation}
with $\Delta_{jk}=\omega_j-\omega_k$. $J_0$ and $\phi_{jk}$ are constants. To understand how the effective magnetic field is derived, let us consider the Hamiltonian of a pair of qubit
\begin{equation}
\hat{H}=\omega\hat{n}_1+(\omega+\Delta)\hat{n}_2+J_0(e^{i\Delta t+i\psi}+e^{-i\Delta-i\phi})(\hat{a}^\dagger_1\hat{a}_2+H.c.)-\frac{U}{2}\sum_{j=1}^2\hat{n}_j(\hat{n}_j-1).
\nonumber
\end{equation}
We then move to a rotating frame via the unitary transformation $e^{-i\Delta \hat{n}_2 t}$. The Hamiltonian in the rotating frame reads
\begin{equation}
\hat{H}=\omega(\hat{n}_1+\hat{n}_2)+J_0(e^{i\Delta t+i\phi}+e^{-i\Delta-i\phi})(\hat{a}^\dagger_1\hat{a}_2e^{-i\Delta t}+H.c.)-\frac{U}{2}\sum_{j=1}^2\hat{n}_j(\hat{n}_j-1).
\nonumber
\end{equation}
If $J_0\ll \Delta$, we can apply the rotating wave approximation and arrive at
\begin{equation}
\hat{H}=\omega(\hat{n}_1+\hat{n}_2)+J_0(\hat{a}^\dagger_1\hat{a}_2e^{-i\phi}+H.c.)-\frac{U}{2}\sum_{j=1}^2\hat{n}_j(\hat{n}_j-1).
\nonumber
\end{equation}
Hence for a three-site lattice with a periodic boundary condition, the total `artifical' magnetic flux is $\Phi=\phi_{12}+\phi_{23}+\phi_{31}$. In the experiment, each site in the superconducting chip is inductively coupled to each other which can be tuned in nanosecond timescale using an external (real) magnetic field. The coupling $J_{jk}(t)/2\pi$ can take any value between -55 MHz to +5 MHz, including zero. The initial state is prepared by applying a $\pi/2$ pulse at the first and the second site.  

\textit{Spectroscopic signatures of localization with interacting photons in superconducting circuits \cite{2018_tangpanitanon_sci}.}  Statistical thermodynamics is one of the pillars of modern physics in describing physical systems with a large degree of freedom. Its fundamental postulate states that all accessible microstates associated with a given macro-state have equal probability. In quantum physics, it has been observed that quantum many-body systems would often evolve and reach a thermal equilibrium over time, regardless of a starting state. However, disorders can prevent those systems from thermalization. The mechanism is known as many-body localization (MBL) \cite{2006_altshuler_ap, 2015_huse_2015,2015_vosk_arcmp,2014_lukin_prl,2010_huse_prb,2014_sondhi_prb,2014_demler_prl, 2014_pollmann_prl,2015_dmitry_prx, 2015_ashvin_natcom, 2016_imbrie_prl, 2016_fazio_prb}. Unlike quantum phase transitions in the equilibrium case such as the Mott to the superfluid phase transition, the thermalized to the MBL phase transition is dynamical and involves all the many-body energy eigenstates of the system. 

Signatures of MBL have been observed in cold atoms in optical lattices \cite{2016_gross_sci,2015_bloch_sci, 2016_markus_sci}, trapped ions \cite{2016_monroe_natphy} and superconducting qubits \cite{PhysRevLett.120.050507}. In all cases, a non-thermal evolution is probed by monitoring the time dynamics of an initially localized state. In the thermalized phase, the system spreads throughout the lattice over time leading to zero population imbalance between sites. When the disorder is increased, and the system is in the MBL phase, the system shows traces of the initial state after a long period. Although this technique reveals signatures of MBL, directly probing many-body energy eigenstates are still absent in the previous work.

In ref. \cite{2018_tangpanitanon_sci}, the authors observe signatures of the celebrated many-body localization transition using interacting photons in a nine-site superconducting circuit. The measurements of the relevant energy eigenenergies and eigenstates were done by implementing a novel many-body spectroscopy method based on time evolution. The approach was benchmarked by measuring the energy spectrum predicted for the system of electrons moving in two dimensions under a strong magnetic field - the Hofstadter butterfly. 

Here we outline the spectroscopy protocol performed in ref. \cite{2018_tangpanitanon_sci}, starting with a single photon. We begin by initializing a photon at site $p$ in the superposition state of $|0\rangle$ and $|1\rangle$, \textit{i.e.},
\begin{align}
|\psi_0\rangle_p &= |0\rangle_1 |0\rangle_2 ...\left(\frac{|0\rangle_p+|1\rangle_p}{\sqrt{2}}\right) ...|0\rangle_{L-1}|0\rangle_{L} \nonumber \\
&= \frac{1}{\sqrt{2}}\left(|\text{Vac}\rangle +| \boldmath{1}_p\rangle\right),  \nonumber \\
&= \frac{2}{\sqrt{2}}\left(|\text{Vac}\rangle +\sum_{\alpha=0}^8 C^p_{\alpha}|E^{(1)}_{\alpha}\rangle\right)
\nonumber
\end{align}
where $| 1_p\rangle = \hat{a}_p^\dagger|\text{Vac}\rangle$, $C^p_{\alpha}=\langle \boldmath{1}_p | E^{(1)}_{\alpha}\rangle$, $|\text{Vac}\rangle$ is the vacuum state, and $|E^{(1)}_{\alpha}\rangle$ is a one-photon energy eigenstate with the eigenenergy $E^{(1)}_\alpha$. The system at time $t$ is given by
\begin{equation}
|\psi (t)\rangle_p = \frac{1}{\sqrt{2}}\left(|\text{Vac}\rangle +\sum_{\alpha}C^p_{\alpha}e^{-iE_{\alpha}^{(1)}t}|E^{(1)}_{\alpha}\rangle\right).
\nonumber
\end{equation}
The operator $\hat{a}_p$ is not Hermitian and therefore not observable. Nevertheless, one can measure 
\begin{align}
\langle \hat{X}_p \rangle &\equiv \langle \hat{a}^\dagger_p + \hat{a}_p\rangle, \nonumber\\
\langle \hat{Y}_p \rangle &\equiv i\langle \hat{a}^\dagger_p - \hat{a}_p\rangle. \nonumber
\end{align}
and compute the expectation value 
\begin{equation}
\langle \hat{a}_p\rangle(t)  \equiv \frac{1}{2}\left(\langle\hat{X}_p\rangle-i\langle\hat{Y}_p\rangle\right) = \frac{1}{2}\sum_{\alpha}|C^{p}_{\alpha}|^2e^{-iE_{\alpha}^{(1)}t}.
\nonumber
\label{eq:mbl:1photon}
\end{equation} 
at different times. The Fourier transform of $\langle \hat{a}_p\rangle(t)$ then reveals the eigenenergies $E^{(1)}_{\alpha}$ and the overlap $|C^p_{\alpha}|$. The experiment is then repeated by varying all possible initial configurations $q\in\{1,2,...,L\}$ and calculating 
\begin{equation}
\chi^{(1)}(t) = \sum_{p=1}^L\langle \hat{a}_p\rangle(t). 
\nonumber
\end{equation}
The latter ensures that all peaks will have appreciable amplitudes in the Fourier transform.

To benchmark the method, the authors implement the 1D Harper model with one photon in a nine-site superconducting circuit, \textit{i.e.},
\begin{equation}
\hat{H}_{\rm Harper}^{\rm 1-photon} = \Delta \sum_{j=0}^8 \cos(2\pi b j )\hat{n}_j  -J\sum_{j=0}^7(\hat{a}^\dagger_j \hat{a}_{j+1}+H.c.).
\nonumber
\end{equation}
The Harper model can be mapped to the 2D Quantum Hall model where $b$ is mapped to a magnetic flux, \cite{2012_kraus_prl}. The spectrum of the Harper model as a function of $b$ exhibits a butterfly-like structure similar to its 2D counterpart. The structure was first proposed by Hofstadter in 1976 \cite{1976_hofstadter_aps}. It is a fractal structure, meaning that small fragments of the structure contain a copy of the entire structure. However, observing the full butterfly in a conventional condensed matter system requires an unphysically large magnetic field in the order of $\sim10^5$ Tesla that can `squeeze' one flux quanta through a unit cell. Signatures of the butterfly were observed by using a superlattice structure in graphene \cite{2013_geim_nat, 2013_kim_nat,2013_ashoori_sci,2013_wolfgag_prl}. 

\begin{figure}
\centering
  \includegraphics[width=0.9\textwidth]{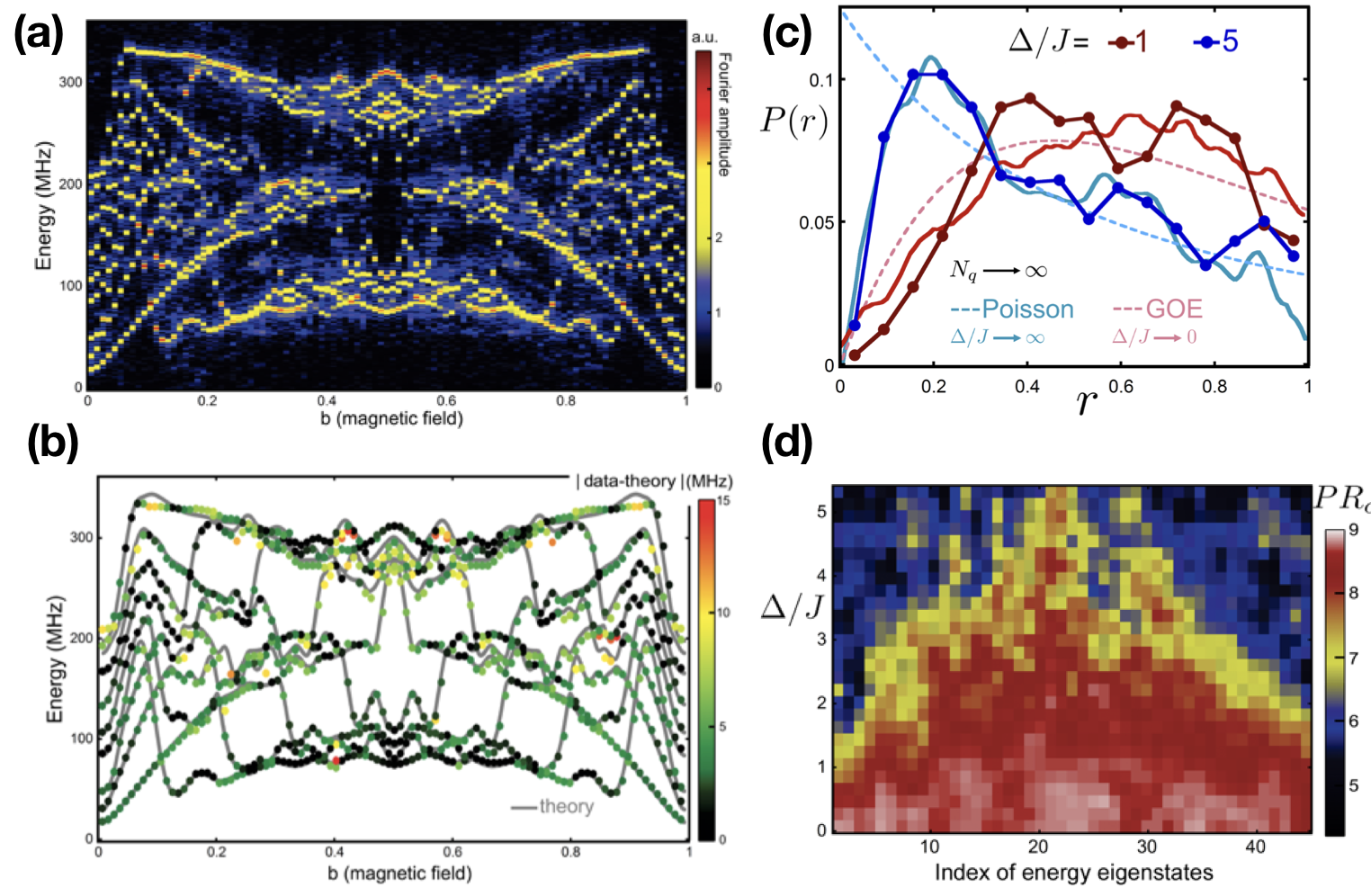}
\caption{\textbf{The Hofstadter butterfly} \textbf{(a)} The Fourier transform of $\chi^{(1)(t)}$ is shown for $100$ values of dimensionless magnetic field $b$ ranging from $0$ to $1$. \textbf{(b)} For each $b$ value, we identify 9 peaks and plot their location as a colored dot.  The numerically computed eigenvalues of the Harper model are shown with gray lines.  The color of each dot is the difference between the measured eigenvalue and the numerically computed one. \textbf{Spectroscopic signatures of localization with two interacting photons}. \textbf{(c)} The measured histogram $P(r)$ of $r_\beta$\,s for $\Delta/J=1$ and 5. The dashed lines are plots of $P_{\text{Poisson}}$ and $P_{\text{GOE}}$ according to eq. (\ref{eq:poi}) and eq.(\ref{eq:goe}), and the solid lines are numerical simulations. \textbf{(d)} The participation ration $PR_{\text{Space}}$ as a function of $\Delta/J$.}
\label{fig:mbl}
\end{figure}

The authors of ref. \cite{2018_tangpanitanon_sci} realize $\hat{H}_{\rm Harper}^{\rm 1-photon}$ by setting the cavity's frequency $\omega_j=\Delta \cos (2\pi b j)$. Since $b$ in $\hat{H}_{\rm Harper}^{\rm 1-photon}$ is not related to a real magnetic, it can be easily tuned from $0$ to $1$ in our setup. Fig. (\ref{fig:mbl}-a) shows the Fourier transform of $\chi^{(2)}(t)$ as a function of $b$. A clear butterfly-like structure is observed as expected since the Fourier transform represents the eigenspectrum. However there are only nine single-particle eigenenergies for each value of $b$, the fractal structure of the spectrum is not displayed. In fig. (\ref{fig:mbl}-b), the measured peaks in the Fourier transform are compared with exact numerics. The errors in the position of the peaks on average are found to be less than $2\%$. This result illustrates high controllability and a low error rate of the setup. 

By placing two photons into the system, the interacting Harper model is then implemented, \textit{i.e.},
\begin{equation}
\hat{H}_{\rm Harper}^{\rm 2-photons} = \Delta\sum_{j=0}^{8} \cos(2\pi bj ) \hat{n}_{j}- \frac{U}{2}\sum_{j=0}^{8} \hat{n}_j(\hat{n}_j-1)-J\sum_{j=0}^7 \left(\hat{a}^{\dagger}_{j+1}\hat{a}_j+\hat{a}^{\dagger}_{j}\hat{a}_{j+1}\right).
\label{eq:2harper}
\nonumber
\end{equation}
Four different irrational values of $b$ are chosen from $\left[0,1\right]$, and the corresponding observables are averaged. The irrational choice of $b$ ensures that the periodicity of the potential and lattice are incommensurate, mimicking the effect of disorder \cite{2015_frahm_epj, 2012_ramaz_epl}. In the experiment, $U/2\pi$ and $J/2\pi$ are fixed to 175 MHz and 50 MHz, respectively ($U/J$ is fixed to 3.5). For this value the ergodic to the localized phase transition of two photons is expected to happen at $\Delta \approx 2J$ \cite{2012_ramaz_epl}.

Spectroscopy of the two-photon eigenstates is done in a similar way, $\textit{i.e.}$, the initial state is prepared as
\begin{align}
|\psi_0\rangle_{p,q} &= |0\rangle_1 |0\rangle_2 ...\left(\frac{|0\rangle_p+|1\rangle_p}{\sqrt{2}}\right) ...\left(\frac{|0\rangle_q+|1\rangle_q}{\sqrt{2}}\right) ...|0\rangle_{L-1}|0\rangle_L  \nonumber \\
&=\frac{1}{2}\left(|\text{Vac}\rangle +|\boldmath{1}_p, \boldmath{1}_q\rangle \right)+\frac{1}{2}\left(| \boldmath{1}_p\rangle+|\boldmath{1}_q\rangle\right),
\label{eq:two_photons_jnit}
\nonumber
\end{align}
where $p\neq q \in\{1,2,..,L\}$ and $|\boldmath{1}_p,\boldmath{1}_q\rangle = |0\rangle_1 |0\rangle_2 ...|1\rangle_p ...|1\rangle_q ...|0\rangle_{L}$ are the two-photon Fock states. Then the measurement at time $t$ reads,
\begin{equation}
\langle\hat{a}_p\hat{a}_q\rangle(t)=\frac{1}{4}\langle\hat{X}_p\hat{X}_q - \hat{Y}_p\hat{Y}_q  - i \hat{X}_p\hat{Y}_q  - i \hat{Y}_p\hat{Y}_q\rangle= \frac{1}{4}\sum_{\beta}|C^{p,q}_{\beta}|^2e^{-iE_{\beta}^{(2)}t}.
\nonumber
\end{equation}
The experiment is then repeated for all possible pairs of $p$ and $q$. The following quantity is then calculated,
\begin{equation}
\chi^{(2)}(t)=\sum_{p\neq q}\langle\hat{a}_p\hat{a}_q\rangle(t)
\nonumber
\end{equation} 
The eigenenergies are then obtained from the Fourier transform of $\chi^{(2)}(t)$.

Since the difference between the ergodic and the localized phase is in its dynamics which are determined by eigenenergies, one of the most direct way to probe the transition is to study the distribution of energy level \cite{2007_huse_prb,2013_roux_prl,1984_schmit_prl}. Using the two-photon protocol, the two-photon eigenenergies $E^{(2)}_{\beta}$ are measured. Then the authors calculate the level spacing $s_{\beta}=E^{(2)}_{\beta+1}-E^{(2)}_{\beta}$ between two adjacent levels and level separation uniformity,
\begin{equation}
r_{\beta} \equiv \frac{\text{min}\{s_{\beta},s_{\beta-1}\}}{\text{max}\{s_{\beta},s_{\beta-1}\}}.
\nonumber
\end{equation}
The level statistics is then defined as a histogram $P(r_{\beta})$ of $r_{\beta}$. In the localized phase when the disorder is large, the levels are uncorrelated resulting in the Poisson distribution
\begin{equation}
 P_{\text{Poisson}}(r_{\beta})=\frac{2}{(1+r_{\beta})^2}.
\label{eq:poi}
\end{equation}
In the ergodic phase, it has been postulated that the statistics of energy levels is the same as the ensemble of real random matrices, following the Gaussian orthogonal ensemble (GOE) \cite{1984_schmit_prl},
\begin{equation}
P_{\text{GOE}}(r_{\beta})=\frac{27}{4} \frac{r_{\beta}+r^2_{\beta}}{(1+r_{\beta}+r^2_{\beta})^{5/2}}.
\label{eq:goe}
\end{equation}

The measured level statistics is shown in fig. (\ref{fig:mbl}-c). It can be seen that at $\Delta < 2J$ the peak of the distribution $P(r)$ is located away from $r=0$. As the disorder is increased beyond $2J$, this peak starts to shift towards $r=0$, as expected from a finite precursor of the thermalized to the MBL phase transition.

The amplitude of the peaks in the Fourier spectrum in the protocol also provides informations about the probability of each energy eigenstate being present at each lattice site $\{P_{\beta,j}\}$. Perhaps, the most common way to quantify the spreading of the eigenstates is to use the participation ratio ($PR$) \cite{2013_huse_prb}
\begin{equation}
PR_{\text{Space}}(\beta)=\frac{1}{\sum\limits_{j} |C_j^{\beta}|^4},
\end{equation}
Here $|C_{\beta}^j|^2$ is the probability of having one or two photons at site $j$. $PR_{\text{Space}}$ indicates the number of lattice sites that are covered by each eigenstate. 

The measured $PR_{\text{Space}}$ as a function of $\Delta/J$ is shown in fig. (\ref{fig:mbl}-d). As $\Delta/J$ is increased, the eigenstates with the highest and the lowest energies start to shrink, and each eigenstate undergoes a delocalized to a localized transition at different disorder strength. This energy-dependent transition is a finite-size signature of the mobility edge in the thermodynamic limit \cite{2016_prb_prb}.


\subsection{Driven-dissipative many-body phases of interacting photons}

Up to now, we have ignored the effect of dissipation by assuming that the dissipation rate is negligibly smaller than a typical energy scale of the system. However, light-matter systems naturally dissipate to the environment. One of the major developments in the field of many-body physics with light is the study of non-equilibrium many-body phases. The latter happens at the steady state where photon losses are compensated by external laser driving. For example, a coupled resonator array as described by the JCH model can be locally driven by a coherent laser field. The total time-dependent Hamiltonian in the lab frame is written as
\begin{equation}
\hat{H}_{\rm tot}(t) = \hat{H}_{\rm JCH} + \sum_{j=0}^{L-1}\Omega_j(\hat{a}^\dagger_j e^{-i\omega_d t} + \hat{a}_je^{i\omega_d t}),
\nonumber
\end{equation}
where $\omega_d$ is the frequency of the drive, $\Omega_j$ is the amplitude of the coherent drive. The time dependence can be removed by going the rotating frame defined by $\hat{U}(t) = \exp \left[i (\sum_{j=0}^{L-1}\hat{a}^\dagger_j\hat{a}_j)\omega_d t\right]$, \textit{i.e.},
\begin{align}
\hat{H}^{\rm R}_{\rm tot} &=  \hat{U}(t)\hat{H}_{\rm tot}(t)\hat{U}(t)^{-1} + i \hat{U}(t)\frac{\partial \hat{U}^{-1}(t)}{\partial t} \nonumber \\
&= \hat{H}_{\rm JCH} -\omega_d \sum_{j=0}^{L-1}\hat{a}^\dagger_j\hat{a}_j+ \sum_{j=0}^{L-1}\Omega_j(\hat{a}^\dagger_j + \hat{a}_j).
\nonumber
\end{align}
The effect of dissipation can be captured by considering the Lindblad master equation
\begin{equation}
\frac{\partial \hat{\rho}}{\partial t} =\mathcal{L}{\hat{\rho}} = -i\left[\hat{H}_{\rm tot}^{\rm R},\rho \right] + \frac{\gamma}{2}\sum_{j=0}^{L-1}\left(2\hat{a}_j\hat{\rho}\hat{a}^\dagger_j-\{\hat{\rho},\hat{a}^\dagger_j\hat{a}_j\}\right)
\label{eq:ness}
\end{equation}
where $\gamma$ is the loss rate, $\hat{\rho}$ is the density matrix of the system, and $\mathcal{L}$ is the Lindblad super-operator. The master equation is then obtained by first writing down the Schr\"odringer for the total system and then tracing out the environment, assuming that the system and the environment are initially in a product state and the bath is memoryless \cite{2008_milburn_springer}. Due to the memoryless bath, the system could reach a non-equilibrium steady state (NESS) that depends on the symmetries of the system, \textit{i.e.},
\begin{equation}
\frac{\partial \hat{\rho}_{\rm NESS}}{\partial t}=0. 
\label{eq:ness}
\end{equation}

Comparison between the NESS of the JCH and the BH model in the driven-dissipative scenario is discussed in ref. \cite{dimitris2012}. Similarities between the two models are found when NESS contains a few photons per site, and the light-matter coupling is much stronger than the dissipation rate $g/\gamma\sim10^4$. In 2009, I. Carusotto's et al. \cite{2009_carusotto_prl} first showed fermionized photons in a driven-dissipative BH array where the NESS of the system mimics a strongly correlated Tonks-Girardeau gas of impenetrable bosons. In an independent work, Hartmann \cite{2010_hartmann_prl} has studied crystallization of photons at the NESS of a similar dissipative BH array but the with alternating local drive $\Omega_j=-\Omega e^{-i\phi_j}$, where $\Omega$ is the amplitude of the drive and $\phi_j=j\pi/2$. Similar behavior is observed in the driven-dissipative JCH array \cite{2012_dimitris_njp}. Signatures of fractional quantum Hall in a 2D driven-dissipative BH array is discussed in \cite{2012_carusotto_prl}. Exotic phases at the NESS includes photon solid phases \cite{2013_jin_prl} and Majorana-like mode of light \cite{2012_bardyn_prl}. The effect of non-linear driving such as parametric down conversion has been discussed in \cite{2012_bardyn_prl, 2013_jonathan_pra}. Probing many-body signatures of non-linear resonator arrays using photon transport have been discussed in ref. \cite{2018arXiv180707882S}.  A nonlinear superconducting circuit with up to 72 sites has also been fabricated to study the dissipative phase transition \cite{2016_houck_prx}. The role of long-range order and the symmetry in driven-dissipative many-body dynamics will be discussed in ref. \cite{jirawat_haldane}.

Below we briefly summarize the main results of some of the above proposals and the experiment done in ref. \cite{2016_houck_prx,2019_schuster_nat}.

\begin{figure}
\centering
\includegraphics[width=0.9\textwidth]{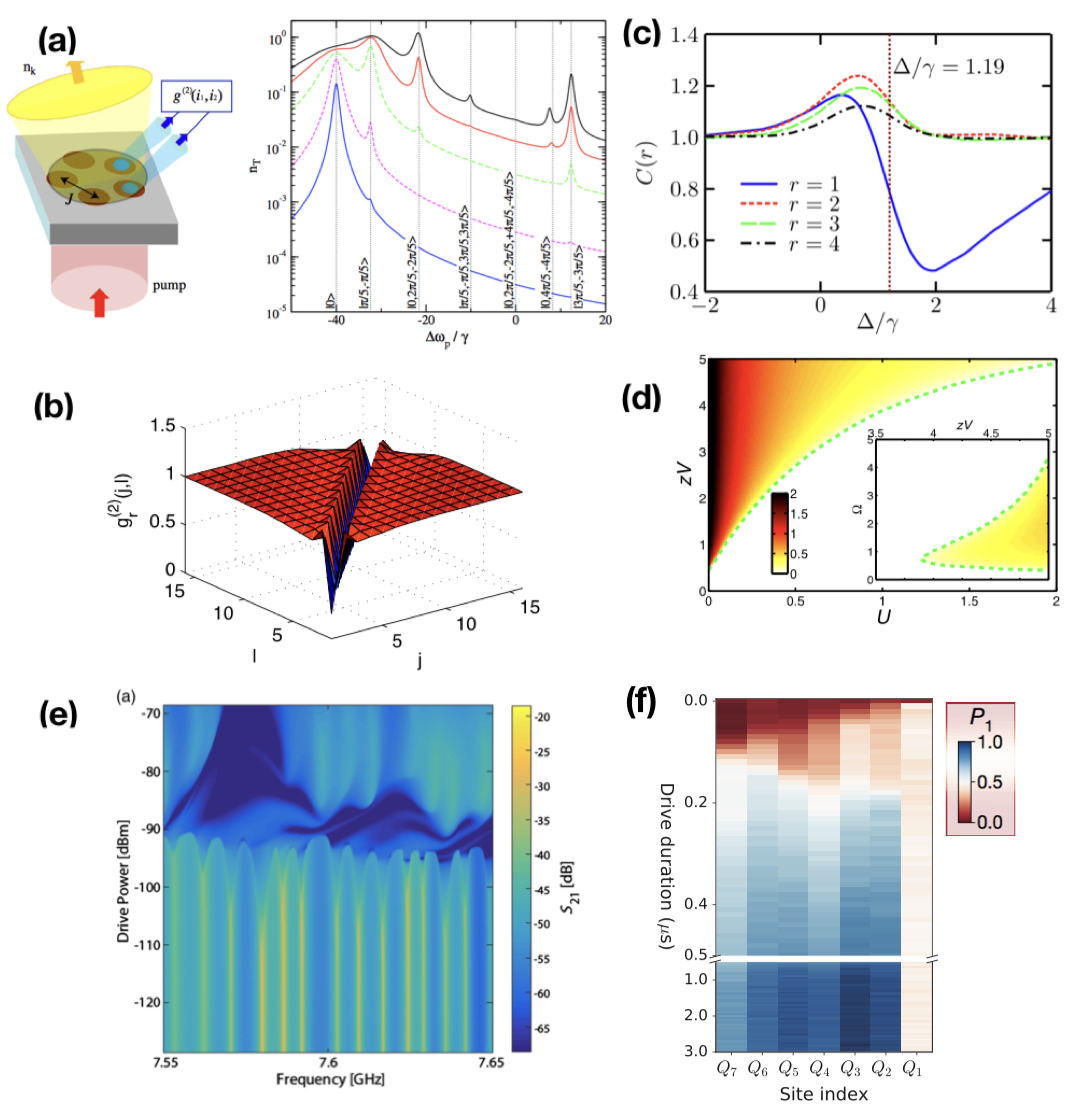}
\caption{\textbf{Driven-dissipative phases and dissipative phase transitions}. \textbf{(a)} fermionized photons at NESS. The plot shows total transmission spectra as a function of the detuning for 5 cavities with $J/\gamma=20$. Difference curves correspond to the pumping amplitude $\Omega/\gamma=0.1,0.3,1,2,3$. \textbf{(b)} Photon crystallization. The figure shows density correlations of the NESS for 16 cavities with $\omega_p=\omega_c$, $U/\gamma=10$, $\Omega/\gamma=2$, and $J/\gamma =2$. \textbf{(c)} Bunching-antibunching transition. Correlations $C(j,r)$ as a function of detuning for $j=30$, $L=61$, $J/\gamma=2$,$\Omega/\gamma=1$. \textbf{(d)} Photon solid phases. The plot shows population imbalance at NESS for zero-detuning, $J=0$, and $\Omega=0.75$. \textbf{(e)} Observation of dissipative phase transitions. The plot shows the transmission as a function of power and driving frequency, exhibiting a transition from a suppressed transmission regime to the regime of dynamical bistability. \textbf{(f)} Dissipatively stabilized Mott Insulator. The plot shows the number of photon, $P_1$, at each site as a function of time.  The results are reproduced from \cite{2009_carusotto_prl, 2010_hartmann_prl,PhysRevA.93.023821,2013_jin_prl,2016_houck_prx,2019_schuster_nat}, respectively.}
\label{fig:dpt}
\end{figure}

\textit{Fermionized photons in an array of driven-dissipative nonlinear cavities \cite{2009_carusotto_prl}.} Here photons blockade is assumed to be strong such that the probability of having two photons at the same site is strongly suppressed. The latter mimics Pauli's exclusion principle of fermionic particles. In this limit, commutation relations of bosonic operators can also be mapped to those of fermionic operators using the Jordan-Wigner transformation. In this work, the authors examine such relation for the steady state of driven-dissipative nonlinear cavities as described by eq. (\ref{eq:ness}). The drive is assumed to be homogeneous. Fig. (\ref{fig:dpt}-a) shows the expectation value of the photon number operator $\langle n\rangle$ at NESS for different value of detuning $\Delta \omega_p=\omega_d-\omega_c$. The peaks happen when the frequency of the drive is resonant with the energy of the non-driven system $E(k) = \omega_c-2J\cos(k)$, where $k$ is the momentum mode. The effect of finite nonlinearities has also been also studied.

\textit{Polariton crystallization in driven arrays of lossy nonlinear resonators  \cite{2010_hartmann_prl}.}
Here the author considers arrays of nonlinear resonators as described by the Bose-Hubbard model. The drive has an alternating phase, $\Omega_j=-\Omega e^{-i\phi_j}$. The correlation function between site $j$ and $l$ at NESS is defined as \begin{equation}
g^{(2)}_r(j,l)=\langle \hat{a}^\dagger_j\hat{a}^\dagger_l\hat{a}_l\hat{a}_j\rangle/\langle\hat{a}^\dagger_j\hat{a}_j\rangle\langle\hat{a}^\dagger_l\hat{a}_l\rangle. 
\nonumber
\end{equation}
For strong nonlinearities,  $g^{(2)}_r(i,j)$ exhibits density-density correlations indicating crystallization of photons, see fig. (\ref{fig:dpt}-b).

\textit{Beyond mean-field bistability in driven-dissipative lattices: Bunching-antibunching
transition and quantum simulation \cite{PhysRevA.93.023821}} 
Here the authors investigated the existence of multiple non-equilibrium states of a driven-dissipative lattice in the limit $U/J\to \infty$. It was found that a commonly-used mean-field approximation which ignores spatial correlations predicts regimes of bistability at the steady state. However, matrix-product-state based analysis reveals that such bistability is an artifact of the mean-field method. The authors also found a bunching-antibunching transition, fig. (\ref{fig:dpt}-c), captured by 
\begin{equation}
C(j,r)=\frac{\langle \hat{n}_j\hat{n}_{j+r}\rangle}{\langle\hat{n}_j\rangle\langle\hat{n}_{j+r}\rangle}, 
\nonumber
\end{equation}
as the detuning $\Delta$ changes.

\textit{Photon solid phases in driven arrays of nonlinearly coupled cavities \cite{2013_jin_prl}}
Here the authors considered arrays described by the extended Bose-Hubbard model with cross-Kerr nonlinearities. The mean-field and the matrix-product-state approaches are used to calculate the NESS phase diagram that includes a photon crystal phase, see fig. (\ref{fig:dpt}-d). The latter is defined by a non-zero population imbalance between two sub-lattices.

\textit{Observation of a dissipative phase transition in a one-dimensional circuit QED lattice \cite{2016_houck_prx}}
In this work, 72 coupled microwave cavities each coupled to a superconducting qubit was fabricated to study dissipative phase transition. Microwave transmission $\langle \hat{a}_j\rangle$ is measured at the NESS. fig. (\ref{fig:dpt}-e) shows the transmission as a function of power and driving frequency, exhibiting a transition from a suppressed transmission regime to the regime of dynamical bistability. 

\textit{A dissipatively stabilized Mott insulator of photons \cite{2019_schuster_nat}} In this work, the authors engineered a dissipative bath to stabilize the Mott insulating phase of photons in a 8-site superconducting circuit. The bath was done by introducing a lossy cavity attached to the system. When a hole in the Mott insulator happens due to a single photon loss, a coherent drive is autonomously applied to replace the hole with two photon on the same size. The extra photon then coherently hops to the lossy cavity and quickly dissipates away. The resulting Mott insulator is therefore dynamically robust again photon loss.


\section{Strongly-interacting photons from superconducting circuits}
\label{circuitQED}
 
\begin{figure}
\centering
\includegraphics[width=0.9\textwidth]{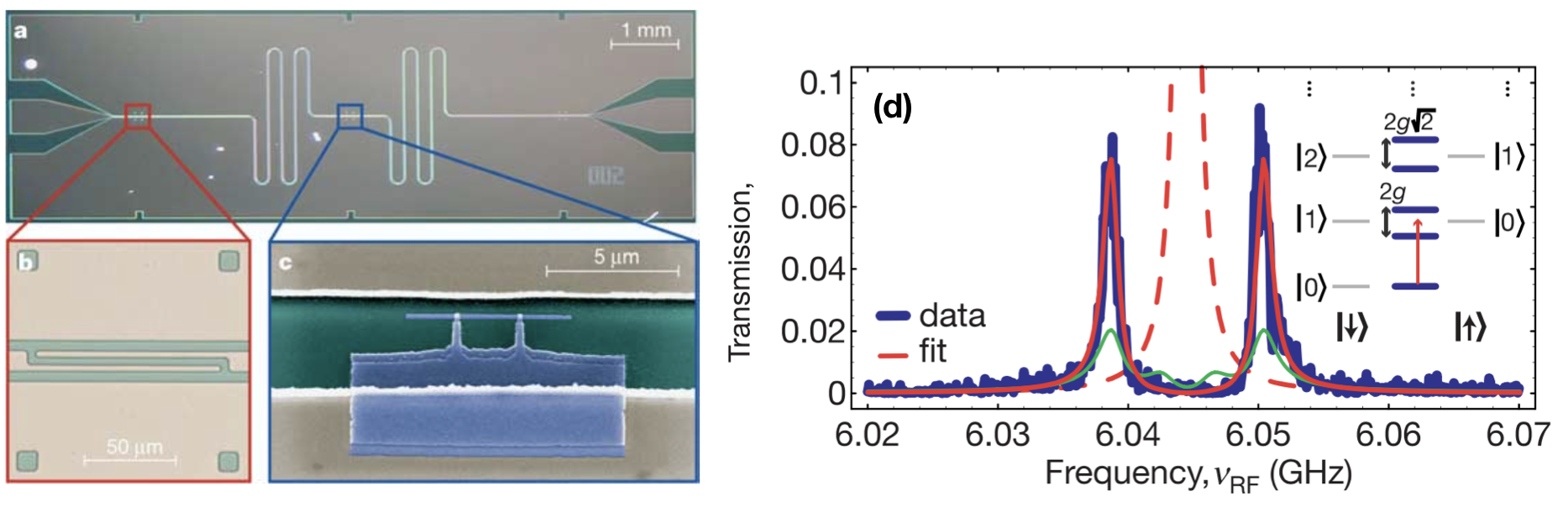}
\caption{\textbf{Strong coupling of a single microwave photon to a superconducting qubit.} \textbf{(a)} The superconducting chip including a coplanar transmission line that act as a cavity. \textbf{(b)} A capacitor to connect the transmission line to an input and output feed. \textbf{(c)} A superconducting qubit that acts as an artificial atom. \textbf{(d)} Vacuum Rabi mode splitting. The results are reproduced from ref. \cite{2004_wallraff_nat}.}
\label{fig:ccqed}
\end{figure}

We now turn our discussion to the implementation of cavity QED and light-matter interactions using superconducting circuits. The idea of quantum phenomena in a macroscopic object is traced back to Josephson in 1962 \cite{JOSEPHSON1962251} who predicted quantum tunneling of Cooper pairs between two superconductors separated by a thin insulating barrier known as a Josephson junction. In the late 1990s, quantized charges and Rabi oscillation of a capacitively shunted Josephson junction subjected to a weak microwave field were observed \cite{1998_devoret_ps,1999_nakamura_nat}, providing evidence that a macroscopic object can behave as an effective quantum two-level system. This `artificial' two-level atom, also known as a superconducting qubit, can then be coupled to modes of a harmonic oscillator such as an LC circuit or a coplanar transmission line. The total system mimics the JC model where a single atom coupled to a cavity. This analogy was put forward in 2004 where strong coupling between a single microwave photon and a superconducting qubit was observed \cite{2004_wallraff_nat}, see fig. (\ref{fig:ccqed}). Since the topology of a circuit can be fabricated almost arbitrarily using the conventional electron-beam lithography, superconducting circuits serves as a scalable platform for quantum simulation with interacting photons. Artificial gauge fields for interacting photons in this system have been realized \cite{2016_roushan_natphy}. A nine-site superconducting circuit with a long coherent time has been fabricated to study signatures of a thermalized to a many-body localized transition \cite{2018_tangpanitanon_sci}. A 72-site superconducting circuit simulating the JCH model has been made to study dissipative phase transition of light as discussed earlier \cite{2016_houck_prx}.

In the following, we will discuss the standard circuit quantization  \cite{1995_devoret} for an LC circuit as a linear element and a particular type of a superconducting qubit called a transmon qubit as a non-linear element. We conclude the section by reviewing state-of-the-art superconducting chips implementing the BH and the JCH model.

\subsection{Microwave photons from an LC circuit}
\label{sec:microwave_photons_from_an_lc_circuit}

\begin{figure}
\centering
\includegraphics[width=0.9\textwidth]{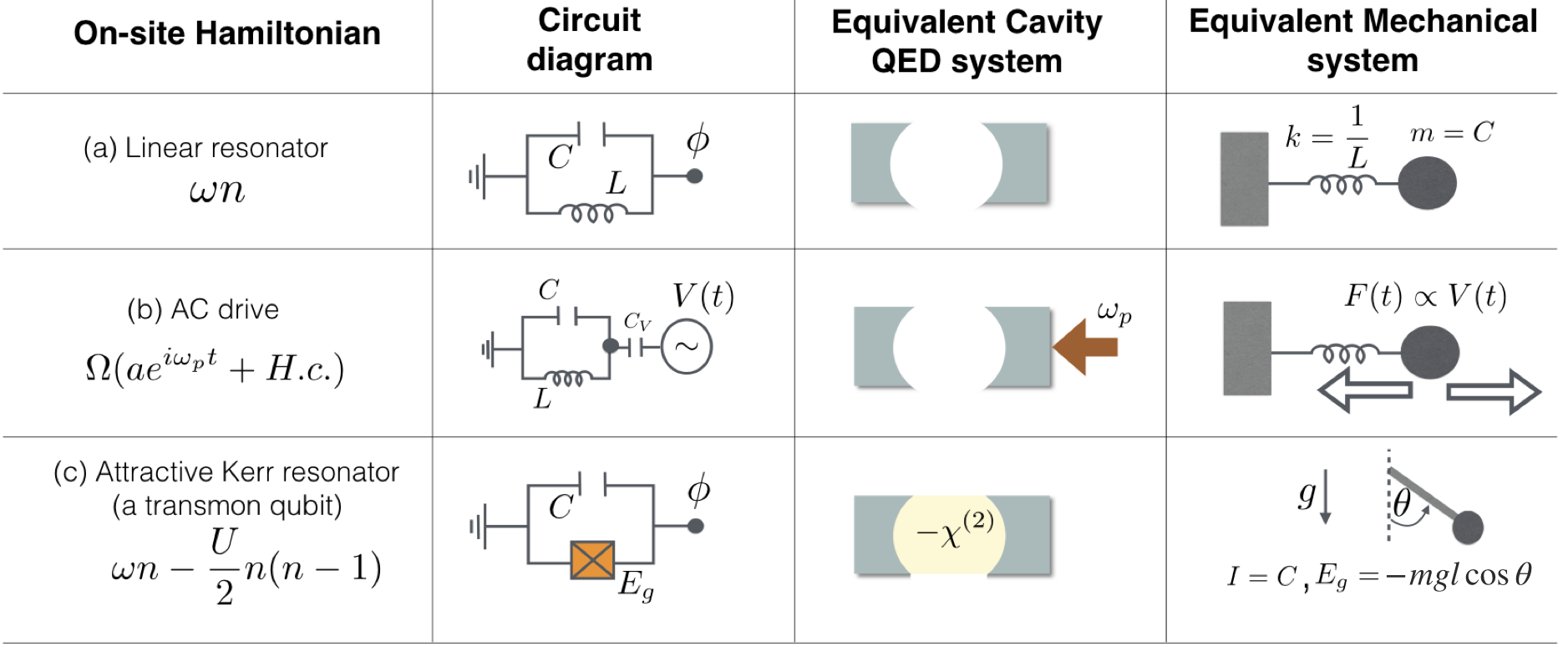}
\caption{\textbf{Basic elements in superconducting circuits.} Because an LC circuit is a harmonic oscillator, it can be viewed as a linear cavity operating in the microwave regime or a mass attached to a spring. The circuit can be driven coherently by applying an external voltage in the same way that an external laser can drive a cavity. A capacitively-shunted Josephson junction (c) behaves like a $\chi^{(2)}$ nonlinear cavity which has an analogy to a mechanical pendulum.}
\label{fig:circuits}
\end{figure}

An LC circuit is depicted in fig. (\ref{fig:circuits}). To write down the Lagrangian for the circuit, we first write down the Kirchhoff's law as  
\begin{equation}
\frac{q}{C}=L\frac{dI}{dt},
\label{eq:lc:kirchhoff}
\end{equation}
where $q$ is the charge stored in the capacitor, $I=dq/dt$ is the current, $C$ is the capacitance, $L$ is the inductance, $\Phi=LI$ is the flux. By differentiating eq. (\ref{eq:lc:kirchhoff}) respect to time, we arrive at an equation of motion for a harmonic oscillator
\begin{equation}
\frac{d^2}{dt^2}q+\omega^2q=0,
\nonumber
\end{equation}
where $\omega = \frac{1}{\sqrt{LC}}$ is the frequency of the oscillator. The energies stored in the capacitor and the inductor are
\begin{equation}
E_C=\frac{q^2}{2C} = \frac{1}{2}C\dot{\Phi}^2
\nonumber
\end{equation}
and 
\begin{equation}
E_L=\frac{1}{2}LI^2 = \frac{1}{2L}\Phi^2,
\nonumber
\end{equation}
respectively, where $\Phi = LI$ is a flux variable. The Lagrangian of the circuit is then defined as
\begin{equation}
\mathcal{L}_{\rm LC} \equiv E_C - E_L = \frac{1}{2}C\dot{\Phi}^2 - \frac{1}{2L}\Phi^2.
\label{eq:lc:kirchhoff}
\end{equation}
This Lagrangian can be compared to that of a particle attached to a spring as shown in fig. (\ref{fig:circuits}-b). The flux $\Phi$ corresponds to the position of the particle $x$. The mass of the particle is $m=C$, while the spring constant is $k=1/L$. Similar to the harmonic oscillator, the Hamiltonian of the LC circuit takes the form
\begin{equation}
H_{\rm LC} \equiv \mathcal{L}_{\rm LC}-\mathcal{Q}\Phi= \frac{\mathcal{Q}^2}{2C} + \frac{\Phi^2}{2L},
\nonumber
\end{equation}
where 
\begin{equation}
\mathcal{Q} = \frac{\partial \mathcal{L}_{\rm LC}}{\partial{\dot{\Phi}}}=C\Phi
\nonumber
\end{equation}
is a conjugate momentum. By promoting $\Phi$ and $Q$ to operators, we get a commutation relation 
\begin{equation} 
 [\hat{Q},\hat{\Phi}]=-i.
 \nonumber
\end{equation}
We define an annihilation operator as
\begin{align}
\hat{a} = i\frac{1}{\sqrt{2C\omega}}\hat{Q}+\frac{1}{\sqrt{2L\omega}}\hat{\Phi}, \nonumber
\end{align}
respectively. The final Hamiltonian then takes the form 
\begin{equation}
\hat{H}_{\rm LC} \approx \hbar\omega(\hat{a}^\dagger \hat{a}+\frac{1}{2}). 
\nonumber
\end{equation}
We note that we choose to discuss an LC circuit here for simplicity. In the experiment, a loss-noise coplanar transmission is usually used as a linear element \cite{2012_houck_pra}. We refer the reader to ref. \cite{1367-2630-14-7-075024} for the detailed derivation of the circuit quantization of the latter.

\subsection{A Kerr resonator from a transmon qubit}
\label{sec:a_transmon_qubit_as_a_large_kerr_nonlinear_resonator}

Josephson junction provides a natural non-linear element for superconducting circuits. A capacitively shunted Josephson junction is described by the Lagrangian 
\begin{equation}
\mathcal{L}_{\rm transmon} = \frac{1}{2}C\dot{\Phi}^2+E_J\cos\left(\frac{\Phi}{\Phi_0}\right),
\label{eq:Ltransmon1}
\end{equation}
where $E_J$ is the Josephson energy and $\Phi_0=\hbar /2e$ is a flux quanta. As shown in fig. (\ref{fig:circuits}), a mechanical analogy of this system is a pendulum where $C$ is the moment of inertia, $E_J$ is the gravitational energy, and $\Phi$ is the angle of the pendulum. Let us first understand the harmonic oscillation limit of this system. Imagine the pendulum is initially placed at its minima $\Phi =0 $ and then subjected to a small kick that generates an oscillation around this point. If the gravitational energy is large compared to the initial kinetic energy, then this oscillation has a small amplitude. In the circuit picture, this corresponds to the limit $E_J/E_C\gg 1$ where $E_C=e^2/2C$. Subsequently, the expansion of the cosine function in $\Phi$ can be truncated at to a finite order, \textit{i.e.},
\begin{equation}
\mathcal{L}_{\rm transmon} \approx \frac{1}{2}C\dot{\Phi}^2-\frac{1}{2L_{J}}\Phi^2 +\frac{E_J}{4\Phi_0^4}\Phi^4+...,
\label{eq:Ltransmon2}
\end{equation}
where 
\begin{equation}
L_J = \Phi_0^2/E_J
\nonumber
\end{equation}
is an effective linear inductance. When keeping up to the second order, the system reduces to a simple Harmonic oscillator with the frequency 
\begin{equation}
\omega=1/\sqrt{L_JC}.
\nonumber
\end{equation}
Slightly away from this limit, the nonlinearity arises from the forth order term. Hence, the system becomes a non-linear oscillator. In the following we will perform the circuit quantization while first keeping infinite orders and only apply truncation after normal ordering of ladder operators. Without truncation, the Hamiltonian can be written as
\begin{equation}
\mathcal{H}_{\rm transmon} = \frac{\mathcal{Q}^2}{2C}+\frac{\Phi^2}{2L_{J}} +\sum_{m=2}^{\infty}\frac{(-1)^mE_J}{(2m)!\Phi_0^{2m}}\Phi^{2m},
\label{eq:Ltransmon2}
\end{equation}
where 
\begin{equation}
\mathcal{Q}=C\frac{\partial \mathcal{L}_{\rm transmon}}{\partial \dot{\Phi}}
\nonumber
\end{equation}
is a conjugate momentum. As before, we promote $\mathcal{Q}$ and $\Phi$ to operators as
\begin{align}
\hat{\Phi} = \sqrt{\frac{L_J\omega}{2}}(\hat{a}^\dagger+\hat{a}), \nonumber \\
\hat{\mathcal{Q}} = i\sqrt{\frac{C\omega}{2}}(\hat{a}^\dagger-\hat{a}),
\end{align}
where $\left[\hat{a},\hat{a}^\dagger\right]=i$. We then apply normal ordering of the operators $\hat{a}$ and $\hat{a}^\dagger$ in $\mathcal{H}_{\rm transmon}$ using the formula \cite{0305-4470-18-2-012}
\begin{equation}
(a+a^\dagger)^{2m}=\sum_{k=0}^{m}\sum_{i=0}^{2m-2k}\frac{(2m)!(a^\dagger )^ia^{2m-2k-i}}{2^kk!i!(2m-2k-i)!}.
\nonumber
\end{equation}
In the limit $E_J/E_C\approx 50 - 100$, the higher order terms in $\mathcal{H}_{\rm transmon}$ can be truncated up to the forth order. As a result, the final Hamiltonian can be written as
\begin{equation}
\hat{H}_{\rm transmon} \approx (\omega+\delta \omega )\hat{n} -\frac{U}{2} \hat{n}(\hat{n}-1),
\label{eq:transmon}
\end{equation}
where $U=-E_Je^{-\lambda^2}\lambda^4/4$ is Kerr nonlinearity,  $\delta \omega = \lambda^2E_Je^{-\lambda^2}$, and $\lambda =(2E_{\tilde{C}}/E_J)^{1/4}$. This Hamiltonian takes the same form of that of a Kerr resonator. Due to the $n$-dependent nonlinearity, a vacuum state $|0\rangle$ and a one-photon Fock state $|1\rangle$ of the resonator can also be used as a qubit. A capacitively shunted Josephson junction operating at this regime is known as a transmon qubit. A typical values of $\omega$ and $U$ are $\sim 5-10$GHz and $\sim 200-300$GHz, respectively  \cite{2009_houck_qip}. Typical lifetime of photons in the transmon qubit is $10-20\mu$s with the dephasing time around $2 \mu$s.

\subsection{Different types of superconducting qubits}

For a larger nonlinearity $E_J/E_{\tilde{C}}>100$, the transmon qubit is also known as a charge or a Cooper-pair-box qubit \cite{1998_devoret_ps,1999_nakamura_nat} which was one of the first qubit design invented in early 1990. However, the charge qubit suffers from charge noises and only has a lifetime of a few $ns$. We note that there are several other designs of superconducting qubits such as flux qubits and phase qubits \cite{2011_Schoelkof_prl, 1999_mazo_prb, 2000_lukens_nat, 2002_martinis_prl, 2009_martinis_qip,2000_mooij_sci,2014_troyer_natphy} for quantum computing applications \cite{2013_schoelkopf_sci}. However, only a transmon qubit can be mapped to a Kerr nonlinear resonator.

\subsection{Nonlinear lattices from arrays of coupled transmon qubits}
\label{sec:many_body_hamiltonians_from_arrays_of_coupled_transmon_qubits}

\begin{figure}
\centering
\includegraphics[width=0.9\textwidth]{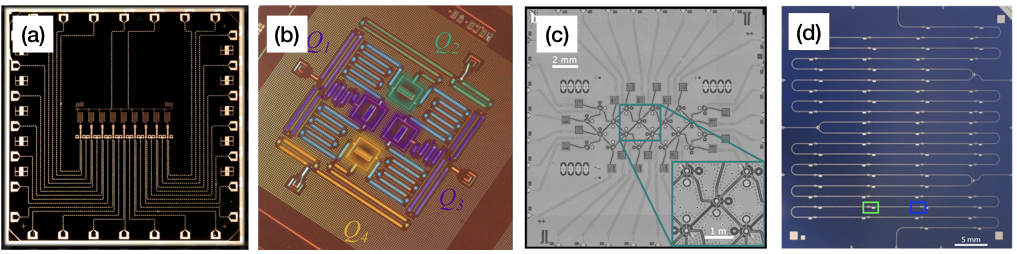}
\caption{\textbf{Arrays of coupled transmon qubits} fabricated by \textbf{(a)} Google with $L=9$ \cite{2018_neill_sci,2018_tangpanitanon_sci} , \textbf{(b)} IBM with $L=5$ \cite{2017_ibm_nat}, \textbf{(c)} Regetti with $L=19$ \cite{2017_regetti_arxiv}. A 72-site superconducting chip implementing the JCH model to study dissipative phase transition \cite{2016_houck_prx} is shown in \textbf{(d)}.}
\label{fig:transmonarray}
\end{figure}

\subsubsection{The Bose-Hubbard model}

Transmon qubits can be coupled in various ways such as a simple use of a capacitor of which we provide details of the circuit quantization below, a transmission line that creates virtual excitation exchange between qubits \cite{2007_raymond_nat,2007_schoelkopf_nat}, and a pair of Josephson junctions that allow the coupling to be tuned in situ using an external flux \cite{2014_martinis_prl, 2015_martinis_pra,2017_hartmut_pra}. Arrays of coupled transmon qubits are described by the Bose-Hubbard Hamiltonian
\begin{equation}
\hat{H}_{\rm BHM}=\sum_{j=0}^{L-1}\omega_j\hat{n}_j - \frac{U}{2}\sum_{j=0}^{L-1} \hat{n}_j(\hat{n}_j-1) -\sum_{\langle j,j'\rangle}J_{j,j'}\left(\hat{a}^\dagger_j\hat{a}_j'+H.c.\right)
\label{eq:bhm}
\end{equation}
where $\hat{n}_j=\hat{a}^\dagger_j \hat{a}_j$ is a local number operator, $J_{j,j'}$ is the hopping coefficient between the sites $j$ and $j'$, $\omega_j$ is the frequency of the resonator $j$. State-of-the-art superconducing chips containing arrays of coupled transmon qubits with different topologies are shown in fig. (\ref{fig:transmonarray}). The 9-site one-dimensional chip in fig. (\ref{fig:transmonarray}-a) was used to implement random circuits for quantum supremacy \cite{2018_neill_sci} and to observe stroboscopic signatures of many-body localization \cite{2018_tangpanitanon_sci}. The 5-site chip in fig. (\ref{fig:transmonarray}-b) and 19-site chip in fig. (\ref{fig:transmonarray}-c) were used to demonstrate quantum variational-based algorithms for quantum chemistry \cite{2017_ibm_nat} and quantum machine learning \cite{2017_regetti_arxiv}.

\begin{figure}
\centering
  \includegraphics[width=0.8\textwidth]{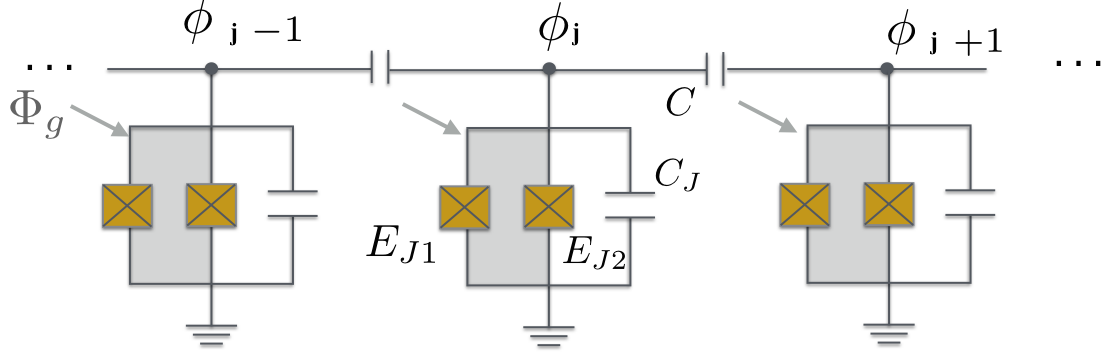}
\caption{Circuit QED diagram showing an implementation of the Bose-Hubbard Hamiltonian. }
\label{fig:thouless:fig5}
\end{figure}

In the following we will give an example of circuit quantization of capacitively-coupled transmon qubits, the circuit diagram is shown in fig. (\ref{fig:thouless:fig5}). The flux variable is defined as $\phi_j=-\int V_j dt$, where $V_j$ is a voltage at the corresponding position. As will be shown below, this quantity can be quantized to the form $\phi_j=\alpha (\hat{a}_j+\hat{a}^\dagger_j)$, where $\hat{a}_j, \hat{a}_j^\dagger$ are bosonic operators at site $j$ and $\alpha$ is some constant that depends on the circuit's elements. As shown in \cite{2007_schoelkof_pra}, two parallel-connected Josephson junction with a flux bias $\Phi_g$ can be thought of as an effective single Josephson junctions $E_J$ where 
\begin{equation}
E_{J}=(E_{J1}+E_{J2})\cos\left(\frac{\Phi_g}{2\Phi_0}\right)\sqrt{1+d^2\tan \left(\frac{\Phi_g}{2\Phi_0}\right)},
\nonumber
\end{equation}
with $\Phi_0=\hbar/2e$ and $d=(E_{J2}-E_{J1})/(E_{J2}+E_{J1})$. The resonator's frequency $\omega_j$ is related to $E_J$, hence it can be tuned on the fly, by changing the flux bias $\Phi_g$.

Following the standard circuit quantization procedure \cite{1995_devoret}, we first write down the circuit's Lagrangian as 
\begin{equation}
\mathcal{L}=\sum_{j=0}^{L-1}\left(\frac{1}{2}C_J\dot{\phi}_j^2+E_J\cos\left(\frac{\phi_j}{\phi_0}\right)\right)+\sum_{j=0}^{L-2}\frac{1}{2}C(\dot{\phi}_j-\dot{\phi}_{j+1})^2,
\nonumber
\end{equation}
Assuming $C/(C_J+2C)\ll 1$, the Hamiltonian can be obtained using the Legendre transformation \cite{Nunnenkamp},
\begin{equation}
H=\sum_{j=0}^{L-1}\left(\frac{\dot{\phi}_j^2}{2\tilde{C}}+\frac{\phi_j^2}{2\tilde{L}}+\sum_{n=2}^{\infty}\frac{(-1)^nE_J}{(2n)!\Phi_0^{2n}}\phi_j^{2n}\right)+\sum_{j=0}^{L-2}\frac{C}{\tilde{C}^2}q_jq_{j+1},
\label{eqApp:H}
\end{equation}
where 
\begin{equation}
q_j=\sqrt{2C+C_J}\frac{\partial \mathcal{L}}{\partial \dot{\phi_j}}
\nonumber
\end{equation}
is a conjugate momentum of $\phi_j$, $\tilde{C}= C_J+2C$ is an effective capacitance and $\tilde{L}=\Phi_0^2/E_J$ is an effective inductance . We then quantized $\phi_j$ and $q_j$ by defining ladder operators $\hat{a}_j$, $\hat{a}^\dagger_j$ according to 
\begin{align}
\hat{\phi}_j  = (\tilde{L}/4\tilde{C})^{1/4}(\hat{a}_j+\hat{a}^\dagger_j)  \\
\hat{q}_j = i(\tilde{C}/4\tilde{L})^{1/4}(-\hat{a}_j+\hat{a}^\dagger_j)
\end{align}
The first two terms in eq. (\ref{eqApp:H}) become $\sum_j\omega \hat{a}^\dagger_j\hat{a}_j$, where $\omega=1/\sqrt{\tilde{L}\tilde{C}}$ is a resonator frequency. In addition, the capacitor $C$ leads to the hopping term with $J=-\frac{\omega C}{2\tilde{C}}$. A rotating-wave approximation is assumed, so we ignore the term $(\hat{a}^\dagger_j\hat{a}^\dagger_{j+1}+h.c.)$.

The Josephson junction $E_J$ introduces an anharmonicity to the resonator's frequency. Due to this anharmonicity, a vacuum state $|0\rangle$ and a one-photon Fock state $|1\rangle$ of the resonator can be used as a qubit. A transmon qubit corresponds to the regime with a large $E_{\tilde{L}}/E_{\tilde{C}}>1$ where $E_{\tilde{C}}=e^2/2\tilde{C}$ and $E_{\tilde{L}}=\Phi_0^2/\tilde{L}$ , such that the terms higher than the forth order can be neglected \cite{2007_schoelkof_pra}. Hence, a transmon qubit can be thought of as a resonator with an attractive Kerr nonlinearity $U<0$. Taking into account the normal ordering \cite{cqed_hartmann}, we get $U=-E_Je^{-\lambda^2}\lambda^4/4$, where $\lambda =(2E_{\tilde{C}}/E_{\tilde{L}})^{1/4}$. This normal ordering also introduces a small normalization factor $\delta \omega$to the resonator frequency, with $\delta \omega = \lambda^2E_Je^{-\lambda^2}$.


\subsubsection{The Jaynes-Cummings Hubbard model}

A single superconducting qubit coupled to a transmission line can be described with the Jaynes-Cummings model, where the transmission line plays a role of a resonator, and a qubit plays a role of an atom. An array of up to 72 coupled Jaynes-Cummings resonators which leads to the JCM model has been implemented in ref. \cite{2016_houck_prx} to study dissipative phase transition. 


\section{Conclusions and future aspects}

Although implementing a universal quantum simulator which requires full control over quantum many-body systems may still be years away, tremendous experimental progress has been made during the past two decades. Two main approaches have emerged. The first approach such cold atom systems provide a global control a large ensemble of quantum particles, with possible local manipulation and measurement in some cases. The second approach such as interacting photons in superconducting circuits provide more flexibility on the local control and measurement while scaling up to 50-100 site are in current progress. For the latter, one needs to develop both new experimental techniques and new theoretical frameworks to maintain such controllability when scaling up. For example, the many-body spectroscopy technique developed in ref. \cite{2018_tangpanitanon_sci} allows one to resolve all energy eigenstates and eigenenergies of the system. This result allows us to benchmark the experiment with the theory and to reconstruct matrix elements of many-body Hamiltonians that a given circuit implements. However, when scaling up eigenenergies of the system will become too dense to be resolved by the current resolution which is limited by the coherence time of the system. Hence, obtaining full information of the Hamiltonian of the circuit is not possible for a large system. To what extent, the disregarded information becomes essential to the physics of the system is still an open question. Constructing the Hamiltonian of the system with limited details also require a new theoretical framework. The latter also raises the question of how to systematically benchmark a quantum simulator as it approaches the limit of classical computers. Identifying problems beyond quantum physics that can only be solved with near-term quantum simulators is also an important question that drives the field forwards. With these in mind, we conclude that, due to exceptional local control systems, interacting photons in superconducting circuits, although still in its early state, is one of the promising candidates for quantum simulation.



\bibliographystyle{iopart-num}
\bibliography{references_list}


\end{document}